\documentclass[aps,superscriptaddress,twocolumn,preprintnumbers,floatfix,showpacs,prx]{revtex4-1}

\usepackage{amsmath,amssymb}
\usepackage{dcolumn}
\usepackage{braket}
\usepackage{dsfont}
\usepackage{feynmp}
\usepackage{hhline}
\usepackage{graphicx}
\usepackage{mathtools}
\usepackage{xfrac}
\usepackage{paralist}
\usepackage{enumitem}
\usepackage[normalem]{ulem}
\usepackage{color}
\definecolor{darkblue}{rgb}{0.,0.,0.4}
\definecolor{darkred}{rgb}{0.5,0.,0.}

\newcolumntype{L}{>{\centering\arraybackslash}m{12.4cm}}

\newcommand{\rb}{\boldsymbol{r}}
\newcommand{\sB}{\boldsymbol{s}}

\newcommand{\beq}{\begin{eqnarray}}
\newcommand{\eeq}{\end{eqnarray}}

\begin{document}

\title{Isotropic Layer Construction and Phase Diagram for Fracton Topological Phases}

\author{Sagar Vijay}
\affiliation{Department of Physics, Massachusetts Institute of Technology,
Cambridge, MA 02139, USA}
\affiliation{Kavli Institute for Theoretical Physics, Santa Barbara, CA 93106, USA}

\begin{abstract}
Starting from an isotropic configuration of intersecting, two-dimensional toric codes, we construct a fracton topological phase introduced in Ref. \cite{Fracton_Gauge_Theory}, which is characterized by immobile, point-like topological excitations (``fractons"), and degenerate ground-states on the torus that are locally indistinguishable. Our proposal leads to a simple description of the fracton excitations and of the ground-state as a ``loop" condensate, 
and provides a basis for building new 3D topological orders such as a natural, $Z_{N}$ generalization of this  fracton phase, which we introduce.  We describe the rich phase structure of our layered $Z_{N}$ system.  By invoking a lattice duality, we demonstrate that when $N \ge 5$, there is an intermediate phase that appears between the decoupled, layered system and the fracton topologically-ordered state, which opens the possibility of a continuous transition into the fracton topological phase.  
We conclude by presenting a solvable model, that interpolates between the fracton phase and a confined phase in which the phase transition is first-order. 
\end{abstract}
\maketitle

Topological phases of matter have been a subject of active interest in condensed matter  physics.  The emergence of local conservation laws in certain quantum many-body systems leads to a convenient, low-energy description of the system as a gauge theory, which captures the behavior of a wide range of topological phases ranging quantum Hall states to quantum spin liquids \cite{Wegner, FradkinShenker, ZhangHassonKivelson, WenNiu, SachdevRead, Wen, SenthilFisher, MoessnerSondhi, Hermele, Wenbook, Halperin, Wilczek, Kitaev_Z2_SL}.  The long-ranged entanglement in the degenerate, locally indistinguishable ground-states of a topological phases have led to proposals for their use as  platforms for universal quantum computation \cite{Kitaev, Nayak, Surf_Code_1, Surf_Code_2, Read}.

In recent years, new kinds of gapped topological phases have been discovered in three spatial dimensions by studying exactly solvable models \cite{Polynomial, Haah, Yoshida, Chamon, Bravyi, Fracton}.   A general feature of these topological phases is the presence of certain point-like topological excitations which are created at the \emph{corners} of operators with support on fractal- or membrane-like regions.  These excitations are strictly immobile at zero temperature, as no local operator can move a single excitations without creating additional gapped excitations in the system.   This is in contrast to the behavior of point-like charges in discrete gauge theories, which may be created at the ends of Wilson lines and behave like mobile particles.  As the point-like excitations in these exotic topological phases appear either as fractions of topological excitations with restricted mobility or at the corners of operators with support on a fractal region of the lattice, they are referred to as ``fractons" \cite{Fracton_Gauge_Theory}.  Apart from providing an exotic alternative to Fermi or Bose statistics in three dimensions, fracton topological orders are of interest for studying clean quantum systems with ``glassy" dynamical behavior \cite{Chamon, Kim_Haah}, and for building robust quantum memories such as Haah's code \cite{Haah, Bravyi_Haah}. The phenomenology of fracton topological phases is thoroughly reviewed in Ref. \cite{Fracton_Gauge_Theory, Fracton}.

Recently, it has been shown that gapped fracton topological phases may be understood through a generalization of conventional lattice gauge theory for interacting quantum systems that have an extensive set of global symmetries (e.g. planar symmetries) \cite{Fracton_Gauge_Theory}.  As a consequence, fracton topological phases may be obtained as the quantum {dual} of these interacting systems.  This provides a generalization of the well-known Wegner duality in (2+1)-dimensions that relates symmetry-breaking phase transitions to the confinement transition of conventional lattice gauge theories \cite{Wegner}.  Furthermore, this generalized lattice gauge theory has been used to find new fracton topological phases of bosons and fermions \cite{Fracton, Fracton_Gauge_Theory}, and to find exotic, interacting systems with fractal symmetries such as the dual of Haah's code \cite{Fracton_Gauge_Theory, Dom}.  Recently, certain gapless phases that have gapped, charge excitations with reduced mobility have also been found in ``higher-rank" $U(1)$ gauge theories \cite{Pretko}.

In this work, we introduce a new construction of a fracton topological phase introduced in Ref. \cite{Fracton_Gauge_Theory} which yields an intriguing connection between two-dimensional topological orders and exotic three-dimensional topological phases that have excitations with reduced mobility.  The starting point for our construction is an isotropic configuration of inter-penetrating, two-dimensional (2D) toric codes \cite{Kitaev}. We demonstrate that condensing appropriate excitations in the layered system -- termed ``composite excitations" -- can lead to (i) {three-dimensional} $Z_{2}$ topological order, (ii) a fracton topological order known as the ``X-cube" phase \cite{Fracton_Gauge_Theory}, whose phenomenology we thoroughly review, or (iii) a topologically trivial paramagnetic state.  Condensing composite excitations effectively binds the charges or fluxes of the two-dimensional toric codes, resulting in an emergent topological excitation with reduced mobility.   Our proposal also gives rise to a ``loop-gas" picture for the ground-state wavefunction of fracton topological phases, as well as a simple understanding of the origin of the immobile excitations, which opens a possible route for constructing other exotic, three-dimensional topological orders.   

We present a natural generalization of this procedure that yields a new, $Z_{N}$ analog of the X-cube phase, before turning our attention to the rich phase structure of this microscopic model. 
 Our results are summarized in the schematic phase diagrams in Fig. \ref{fig:Phase_Diagram}a and b. Of importance is a lattice duality between certain phase transition(s) into the fracton topological phase and a confinement transition in a conventional (3+1)-d $Z_{N}$ gauge theory driven by the condensation of $Z_{N}$ flux loops.  From this duality and from the well-known phase structure of $Z_{N}$ gauge theories, we determine that when $N \ge 5$, there must be an intermediate phase between the decoupled layers and the fracton topological phase, with an emergent $U(1)$ gauge symmetry. While we believe this phase is generically gapped and topologically trivial, we determine that this phase remains gapless along a specific line of the phase diagram, along which the phase transition into the fracton topological phase is believed to be continuous.  This opens the possibility of a continuum field theory description of this particular fracton topological phase, which we leave to future work \cite{SV_new} to explore.  We conclude by presenting a solvable projector model in which the transition from the fracton topological phase to a trivial confined phase is first order, though this may not capture the generic behavior of this transition.  


 \begin{figure}
 \includegraphics[trim = 0 0 0 0, clip = true, width=0.5\textwidth, angle = 0.]{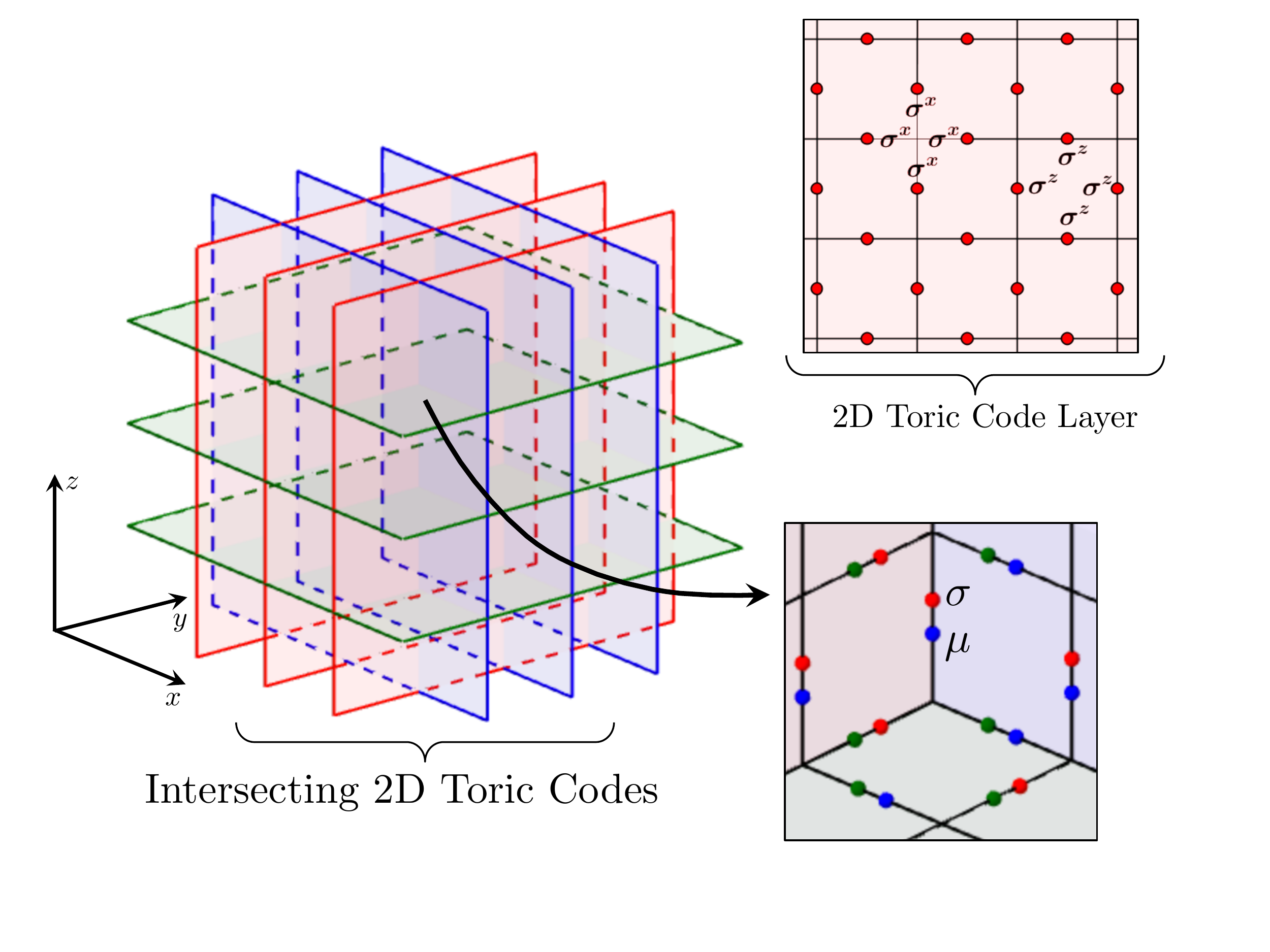}
 \caption{{\bf Intersecting Layers of 2D Toric Codes}: A stack of two-dimensional, square-lattice toric codes in the $xy$ (green), $yz$ (red) and $xz$ (blue) directions, which intersect at sites.  The resulting three-dimensional cubic lattice has two spins per \emph{link} ($\sigma$, $\mu$) as shown.  A single layer of the square-lattice toric code is shown as well, with the ``star" and ``plaquette" operators defined as shown.}
  \label{fig:3D_Stack}
\end{figure}

\section{Isotropic Layer Construction}
We begin by introducing the microscopic model for our layer construction, before analyzing its rich phase structure and the emergence of fracton topological order.  Consider a single layer of the two-dimensional (2D) toric code on the square lattice \cite{Kitaev}, which describes the zero-correlation length limit of the deconfined phase of two-dimensional $Z_{2}$ gauge theory.  The $Z_{2}$ gauge field lies on the links of the square lattice, and at each lattice site and plaquette, there are four-spin operators that measure the $Z_{2}$ charge and flux, respectively (the so-called ``star" and ``plaquette" operators), as shown in the inset in Fig. \ref{fig:3D_Stack}.  We take as our convention that a star operator is given as the product of the $x$-component of the Pauli spins, while a plaquette operator involves a product of the $z$-component.
 
We now place $L$ copies of the square-lattice toric code in the $xy$, $yz$, and $xz$ planes, respectively; any three mutually orthogonal planes intersect at a single {site} of the 2D square lattice. As shown schematically in Fig. \ref{fig:3D_Stack}, this intersecting, three-dimensional arrangement of the toric codes forms a three-dimensional (3D) cubic lattice with $L^{3}$ sites and \emph{two} spins per link.  The 2D $Z_{2}$ charge operators lie at the sites of the cubic lattice while the $Z_{2}$ flux operators lie at plaquettes. Initially, different copies of the 2D toric codes are completely decoupled, and the Hamiltonian for the full system in terms of the two species of spins ($\sigma$ and $\mu$) at each link may be written in the form

 \begin{figure}
 $\begin{array}{ccc}
\includegraphics[trim = 0 0 0 0, clip = true, width=0.2\textwidth, angle = 0.]{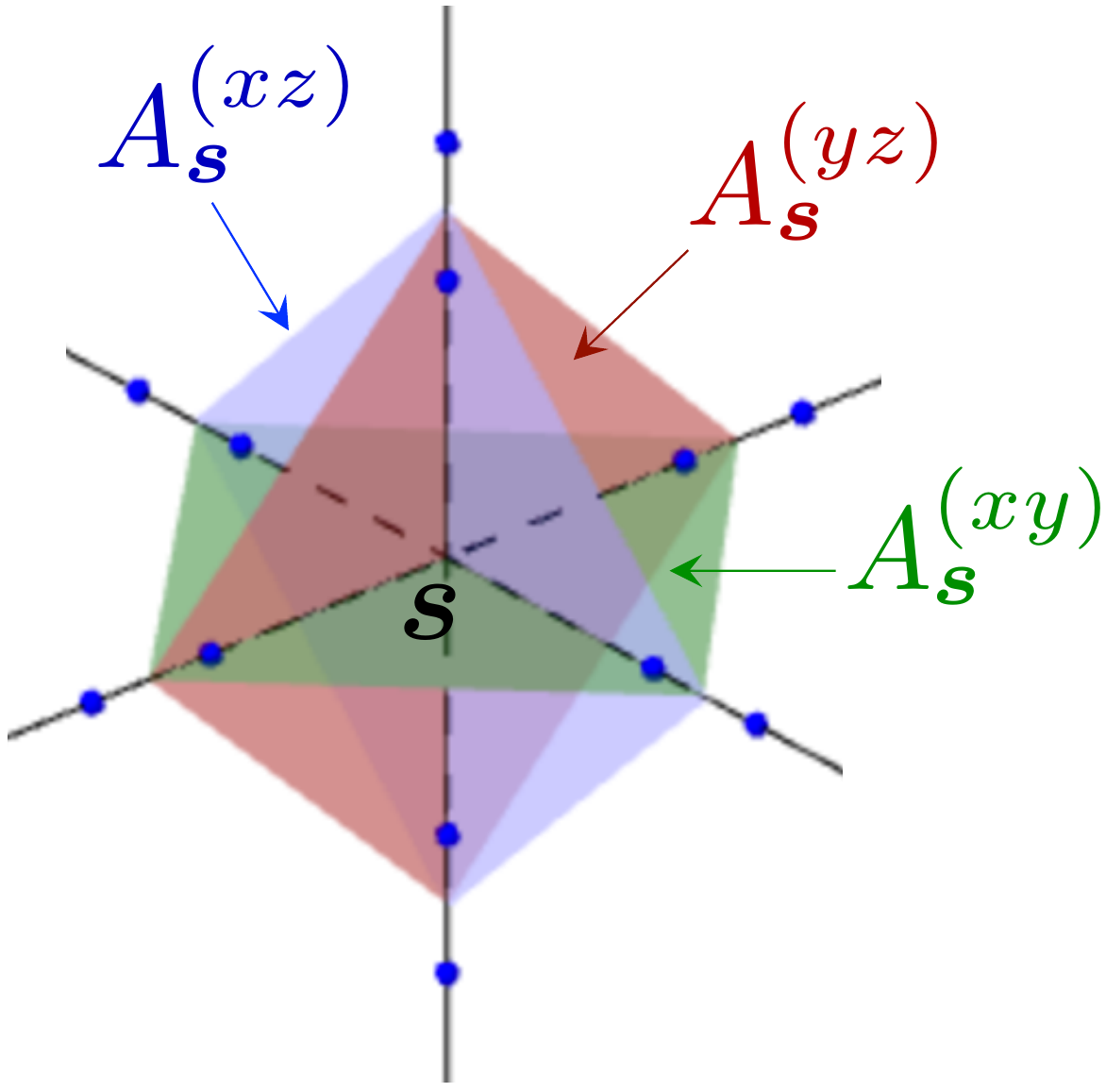} &
 & \includegraphics[trim = 0 0 0 0, clip = true, width=.21\textwidth, angle = 0.]{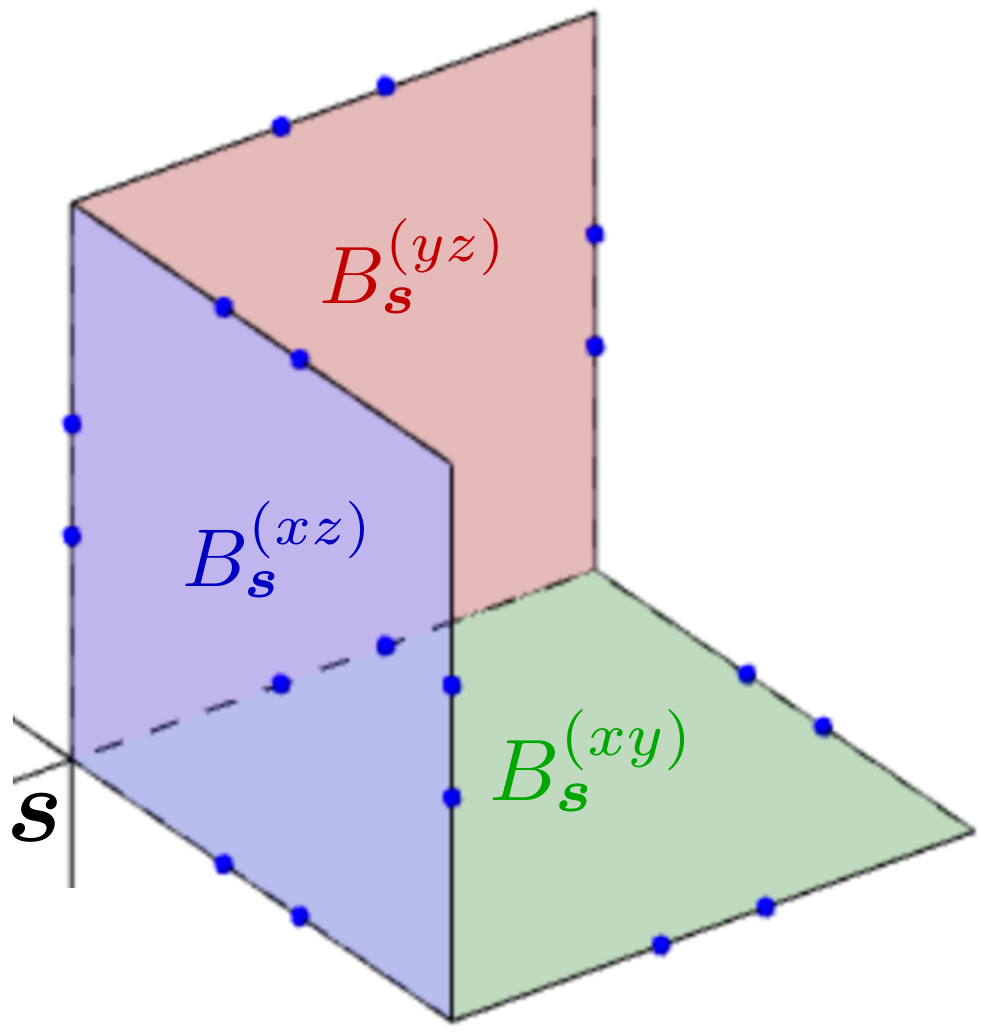}\\
 & & \\
\text{(a)\hspace{.3in}} & & \text{(b)}
 \end{array}$
 \caption{{\bf $Z_{2}$ charge \& flux operators for the decoupled layers}: The locations of the (a) $Z_{2}$ charge and (b) $Z_{2}$ flux operators at each site of the three-dimensional cubic lattice for the decoupled, intersecting layers of two-dimensional toric codes; each charge or flux operator is oriented along the $xy$ (green), $yz$ (red), or $xz$ (blue) planes. 
 }
  \label{fig:Decoupled_Operators}
\end{figure}

\begin{align}
H_{0} = -J\sum_{\boldsymbol{r}}\sum_{ j\,=\,xy,\,yz,\,xz }\left[A_{\boldsymbol{r}}^{(j)} + \,B_{\boldsymbol{r}}^{(j)}\right]
\end{align}
where the sum is over the sites $\boldsymbol{r}$ on the cubic lattice.
Here, $A_{\boldsymbol{r}}^{(j)}$ is the four-spin star operator at site $\rb$ which measures the $Z_{2}$ charge along the copy of the square-lattice toric code that is oriented along the $j^{\mathrm{th}}$  plane (with $j = xy$, $yz$, or $xz$).  
We associate three plaquette operators $B_{\boldsymbol{r}}^{(j)}$ with each site of the cubic lattice, which measure the elementary $Z_{2}$ flux through an elementary plaquette oriented in the $j^{\mathrm{th}}$ plane.  The star and plaquette operators in the 3D cubic lattice are shown schematically in Fig. \ref{fig:Decoupled_Operators}.   All of the operators in $H_{\mathrm{toric}}$ mutually commute, and the ground-state may be written exactly as an equal-amplitude superposition of closed electric-charge loops within each plane.  The topological degeneracy $D$ is simply $\log_{2}D = 6L$ after imposing periodic boundary conditions.  The topologically-degenerate ground-states are locally indistinguishable, so that the 2D $Z_{2}$ topological order of the decoupled toric codes is stable in the presence of weak perturbations \cite{BHM}.


We now add an interaction 
\begin{align}\label{eq:Decoupled}
H_{1} = - h\sum_{\langle \boldsymbol{r},\boldsymbol{r}'\rangle}\sigma^{z}_{\boldsymbol{r}\boldsymbol{r}'}\mu^{z}_{\boldsymbol{r}\boldsymbol{r}'} -t\sum_{\langle \boldsymbol{r},\boldsymbol{r}'\rangle}\sigma^{x}_{\boldsymbol{r}\boldsymbol{r}'}\mu^{x}_{\boldsymbol{r}\boldsymbol{r}'} 
\end{align}
which couples the two spins on each link of the cubic lattice.  We begin by analyzing the  phase diagram of the Hamiltonian 
\begin{align} \label{eq:H}
H = H_{0} + H_{1}.
\end{align}
The ground-state of (\ref{eq:H}) will have the same topological order as a decoupled set of $3L$ toric codes when $t$, $h\ll J$.  We refer to this as the ``decoupled phase" for the remainder of this work. In addition, however, we will demonstrate that this Hamiltonian realizes (i) a (3+1)-d $Z_{2}$ topological phase and (ii) a fracton topological phase, along with (iii) a trivial paramagnet which corresponds to a confined limit of the two topological phases.

 \begin{figure}
 \includegraphics[trim = 0 0 0 0, clip = true, width=0.43\textwidth, angle = 0.]{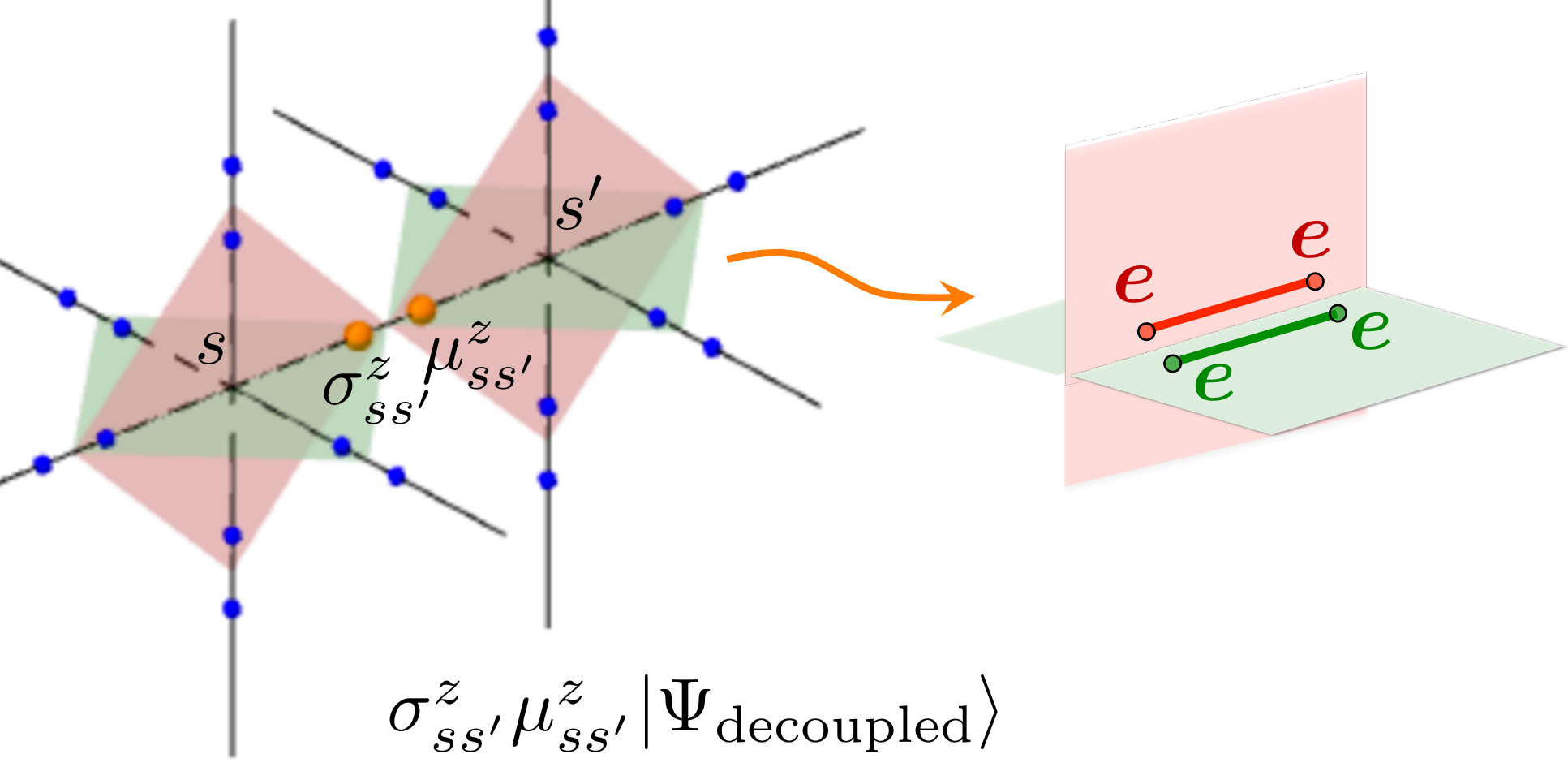}
 \caption{{\bf Composite Charge Condensation}: The operator $\sigma^{z}_{\sB\sB'}\mu^{z}_{\sB\sB'}$, when acting on the ground-state $\ket{\Psi_{\mathrm{decoupled}}}$ of the decoupled, two-dimensional toric codes, creates four electric charge excitations, two in each of the orthogonal planes that meet at the link, as shown.  Condensing this composite charge excitation leads to a three-dimensional $Z_{2}$ topological phase. }
  \label{fig:Condensation_Z2}
\end{figure}

\subsection{(3+1)-d $Z_{2}$ Topological Order from Composite Charge Condensation}
Starting from the decoupled phase, we now increase $h$, while keeping $J$ and $t$ fixed, which eventually leads to a \emph{$(3+1)$}-dimensional $Z_{2}$ topological phase.  Before demonstrating this explicitly with the microscopic Hamiltonian (\ref{eq:H}), we obtain intuition for this result by studying how the ground-state wavefunction changes as $h$ is increased.  When acting on the decoupled toric codes, the operator  $\sigma^{z}_{\boldsymbol{s}\boldsymbol{s}'}\mu^{z}_{\boldsymbol{s}\boldsymbol{s}'}$ creates four electric charge excitations in the orthogonal layers that meet at the link connecting sites $\sB$ and $\sB'$, as shown in Fig. \ref{fig:Condensation_Z2}.  Increasing $h$ leads to the condensation of this excitation, which we refer to as the \emph{composite electric charge}. 

\begin{figure}
 \includegraphics[trim = 0 0 0 0, clip = true, width=0.48\textwidth, angle = 0.]{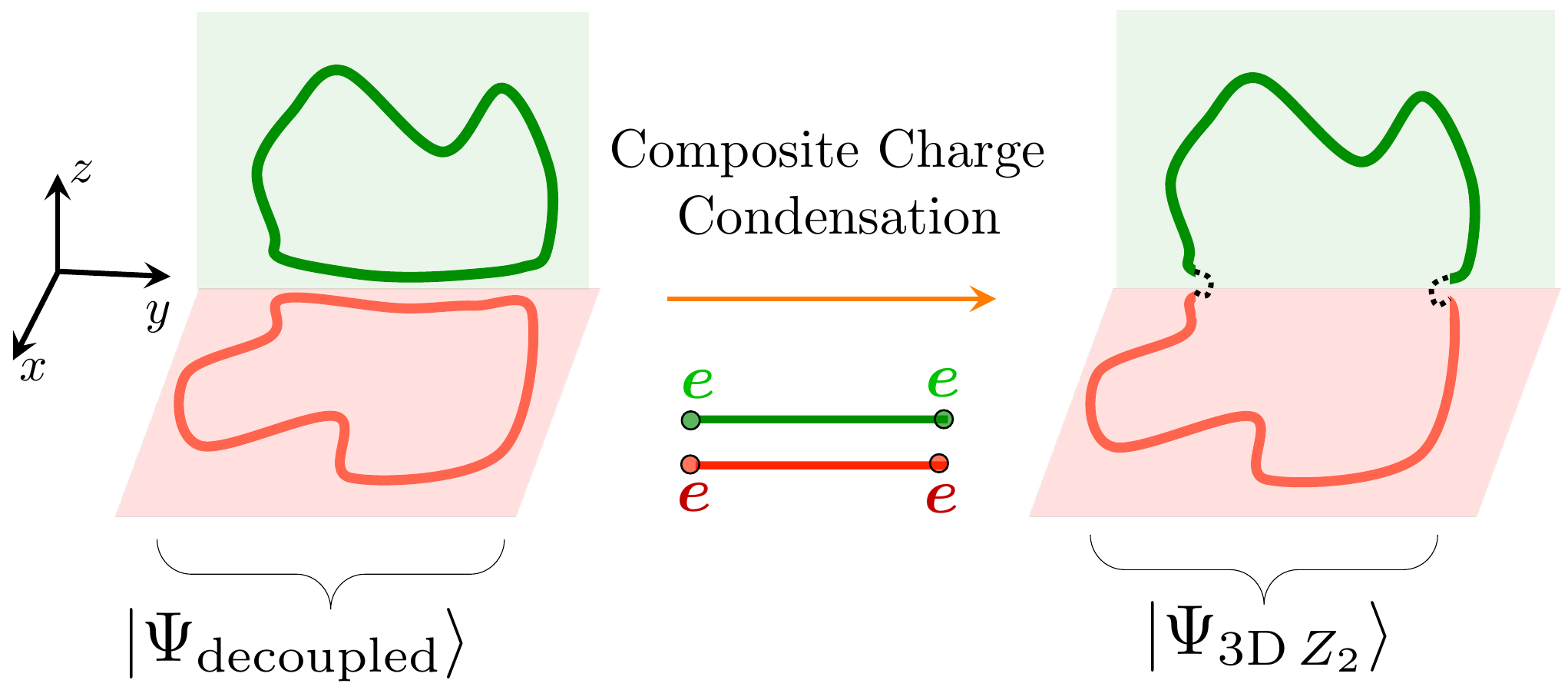} 
  \caption{{\bf ``Gluing" Loops}: Condensing the composite charge excitation has the effect of ``gluing" electric charge loops in adjacent, orthogonal layers.  The resulting wavefunction is that of the 3D $Z_{2}$ topological phase, as described in Sec. I. }
  \label{fig:Loop_Wavefunction}
\end{figure}

Condensing the composite electric charge ``glues" the planar electric charge loops in the wavefunction for the decoupled toric codes ($\ket{\Psi_{\mathrm{decoupled}}}$). As shown in Fig. \ref{fig:Loop_Wavefunction}, adding a composite electric charge has the effect of ``cutting open" two electric charge loops in orthogonal layers that appear in a loop configuration in the state $\ket{\Psi_{\mathrm{decoupled}}}$. 
This ``cut" loop configuration is to be interpreted as a {single}, three-dimensional loop of the \emph{emergent} electric charge $A_{\sB}^{(xy)}A_{\sB}^{(yz)}A_{\sB}^{(xz)}$, which remains well-defined in the condensed phase, and the resulting loop superposition is precisely the wavefunction for the 3D $Z_{2}$ topological phase.  As we will soon show, composite charge condensation also has the effect of binding the 2D $Z_{2}$ fluxes into flux loops.  In this way, the 3D electric charge inherits its bosonic self-statistics and $\pi$-mutual statistics with flux loops from the 2D $Z_{2}$ phase.   

We now derive this result from the microscopic Hamiltonian (\ref{eq:H}).  When $h \gg J$, $t$, we let $\sigma_{\rb\rb'}^{z}\mu_{\rb\rb'}^{z} = +1$ and define a single spin degree of freedom for every link of the cubic lattice as
\begin{align}
\tau^{z}_{\rb\rb'} \equiv \sigma_{\rb\rb'}^{z} \hspace{.4in}  \tau^{x}_{\rb\rb'} \equiv \sigma_{\rb\rb'}^{x}\mu_{\rb\rb'}^{x}
\end{align}  
In terms of this spin, we obtain an effective Hamiltonian in perturbation theory
\begin{align}
H_{\mathrm{eff}}^{(1)} = -\widetilde{J}\,\sum_{\rb}\widetilde{A}_{\rb} - J\sum_{p}B_{p} - t\sum_{\langle\rb,\rb'\rangle}\tau^{x}_{\rb\rb'}
\end{align}
where the coupling $\widetilde{J} \sim O(J^{3}/h^{2})$.  Here, the operator
\begin{align}
\widetilde{A}_{\rb} \equiv A^{(xy)}_{\rb}A^{(yz)}_{\rb}A^{(xz)}_{\rb} = \prod_{\rb'\in\mathrm{star}(\rb)}\tau^{x}_{\rb\rb'}
\end{align}
is precisely the six-spin operator that measures the $Z_{2}$ charge in a {(3+1)}-dimensional $Z_{2}$ topological phase, while
\begin{align}
B_{p} \equiv \prod_{\rb,\rb'\in \partial p}\tau^{z}_{\rb\rb'}
\end{align}
is the four-spin operator measuring the flux through plaquette $p$ on the cubic lattice.  When $t = 0$, the effective Hamiltonian $H_{\mathrm{eff}}^{(1)}$ is precisely that of the 3D toric code \cite{Hamma}. Furthermore, increasing $t$ eventually leads to condensation of the $Z_{2}$ flux loops, resulting in a trivial, confined phase.  As advertised, the 2D $Z_{2}$ fluxes are confined, while a {bound-state} of the $Z_{2}$ fluxes survives as a topological excitation, which becomes the  flux loop of the three-dimensional $Z_{2}$ topological phase.  Condensing a composite excitation has led to the emergence of a new topological excitation whose mobility is restricted.  



\subsection{``{X-Cube}" Fracton Topological Order from Composite Flux Loop Condensation} 
While composite charge condensation leads to three-dimensional $Z_{2}$ topological order, condensing a bound-state of the $Z_{2}$ fluxes in adjacent layers -- a \emph{composite {flux loop}} -- yields the so-called ``X-cube" fracton topological phase, as originally introduced and studied in Ref. \cite{Fracton_Gauge_Theory} using an exactly solvable, commuting Hamiltonian.  We first review the phenomenology of the X-cube fracton topological phase before demonstrating that this phase emerges within our layer construction.  A detailed description of this fracton phase is provided in Ref. \cite{Fracton_Gauge_Theory}.

\subsubsection{Phenomenology of the X-cube Phase}

The X-cube phase is a gapped topological phase that was introduced in Ref.  \cite{Fracton_Gauge_Theory}, by studying the dual description of  the plaquette Ising model, an interacting spin system with an extensive set of planar spin-flip symmetries; the finite-temperature behavior of this system has been studied previously \cite{Savvidy_Wegner, Savvidy_2, Savvidy_3, Wegner_2}. While the plaquette Ising model exhibits only symmetry-breaking or paramagnetic phases at zero temperature,  its dual description is far more exotic, describing the ``confinement'' transition of a fracton topological phase.  The solvable Hamiltonian for this fracton phase, termed the X-cube model, due to the geometry of the multi-spin interactions \cite{Fracton_Gauge_Theory}, was shown to have a  topological ground-state degeneracy $\log_{2}D = 6L-3$ on the length-$L$ three-torus, along with exotic topological excitations whose mobility is severely restricted.  Since the degenerate ground-states are locally indistinguishable, this degeneracy is stable to perturbations, so that the X-cube model \cite{Fracton_Gauge_Theory} describes a stable, gapped {phase} of matter.  

The X-cube phase has two types of gapped, topological excitations.  The first are point-like, immobile excitations (the ``fractons") which are created by acting on the ground-state with an operator supported on a flat, rectangular region.  The fracton excitations are created at the \emph{corners} of this membrane and may only be created in groups of four.  No local operator can move a single fracton  without creating other excitations in the system. While individual fracton excitations are {immobile}, a \emph{pair} of fractons are mobile within a plane and have bosonic self-statistics.  In the X-cube fracton phase, there is a second type of point-like topological excitation, referred to as a ``dimension-1 quasiparticle", which may only move along straight lines without creating additional excitations.  Pairs of fractons which are mobile within planes exhibit $\pi$ mutual statistics with dim.-1 quasiparticles contained within their plane of motion.   

\subsubsection{Isotropic Layer Construction}
\begin{figure}
 \includegraphics[trim = 0 0 0 0, clip = true, width=0.43\textwidth, angle = 0.]{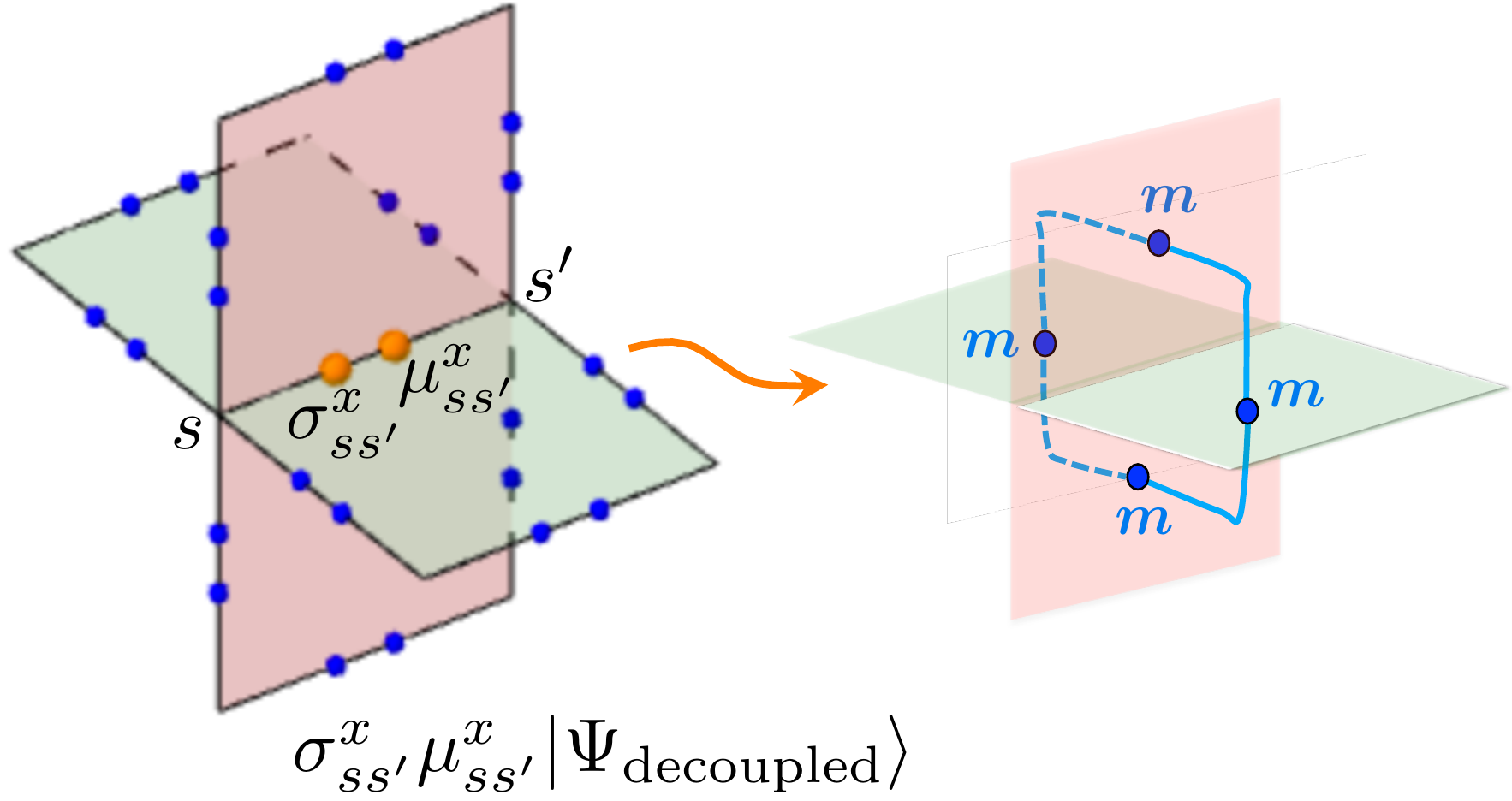} 
 \caption{{\bf Composite Flux Loop Condensation}: The operator $\sigma^{x}_{\sB\sB'}\mu^{x}_{\sB\sB'}$, when acting on the decoupled layers of the two-dimensional $Z_{2}$ phase, creates four flux excitations in orthogonal planes, as shown.  Condensing this composite flux leads to the ``X-cube" fracton topological phase, with fracton operator $\mathcal{O}_{c}$ as given in the main text, and shown in Fig. \ref{fig:Z_N_Xcube}.}
  \label{fig:Condensation_Fracton}
\end{figure}

 To observe the emergence of the X-cube phase in our layer construction, we start from the decoupled phase of the Hamiltonian (\ref{eq:H}) and increase $t$, while keeping $J$ and $h$ fixed.  Acting with the operator $\sigma^{x}_{\rb\rb'}\mu^{x}_{\rb\rb'}$ on the ground-state of the decoupled planes of toric codes creates four $Z_{2}$ fluxes in orthogonal layers, which form a closed loop on the dual lattice as shown in Fig. \ref{fig:Condensation_Fracton}a. We refer to this excitation as a \emph{composite flux loop}.  

The X-cube fracton phase emerges after condensation of this composite flux loop.  When $t \gg J$, $h$, we let $\sigma_{\rb\rb'}^{x}\mu_{\rb\rb'}^{x} = +1$ and define an effective spin degree of freedom for every link of the cubic lattice as $\tau^{x}_{\rb\rb'} \equiv \sigma^{x}_{\rb\rb'}$, $\tau^{z}_{\rb\rb'} \equiv \sigma_{\rb\rb'}^{z}\mu_{\rb\rb'}^{z}$.  We then obtain an effective Hamiltonian in perturbation theory as
\begin{align}
H_{\mathrm{eff}}^{(2)} = H_{\mathrm{X}\text{-}\mathrm{cube}} - h\sum_{\langle\rb,\,\rb'\rangle}\tau^{z}_{\rb\rb'}
\end{align}
where
\begin{align}
H_{\mathrm{X}\text{-}\mathrm{cube}} = -J\sum_{\rb,j}A_{\rb}^{(j)} - K\sum_{\rb}\mathcal{O}_{\rb}
\end{align}
with $K \sim O(J^{6}/t^{5})$.  Here, $A_{\rb}^{(j)}$ is a four-spin operator for the spins along the four links emanating from site $\rb$ that lie in the $j^{\mathrm{th}}$ plane:
\begin{align}
A_{\rb}^{(j)} = \prod_{\sB\in\mathrm{plane}_{j}(\rb)}\tau^{x}_{\rb\sB}
\end{align}
while the operator $\mathcal{O}_{\rb}$, given by
\begin{align}\label{eq:Cube_Op}
\mathcal{O}_{\rb} &\equiv B_{\rb}^{(xy)}B_{\rb}^{(yz)}B_{\rb}^{(xz)}B_{\rb + \hat{x}}^{(yz)}B_{\rb + \hat{y}}^{(xz)}B_{\rb + \hat{z}}^{(xy)}\nonumber\\
 &= \prod_{\rb,\rb'\in\mathrm{cube}(\rb)}\tau^{z}_{\rb\rb'}
\end{align}
is precisely the product of the twelve $\tau^{z}$ spins surrounding an elementary cube as shown in Fig. \ref{fig:Z_N_Xcube}.  The Hamiltonian $H_{\mathrm{X-cube}}$ is precisely the commuting Hamiltonian that describes the ``fixed-point" properties of the X-cube fracton phase, as introduced in Ref. \cite{Fracton_Gauge_Theory}. The ground-state at the solvable point satisfies $\mathcal{O}_{\rb}\ket{\Psi} = A_{\rb}^{(j)}\ket{\Psi} = \ket{\Psi}$ for all $\rb$, $j$.  Excitations may be created by acting with string or membrane operators to violate these constraints. The operators $\mathcal{O}_{\rb}$ and $A_{\rb}^{(j)}$ measure the $Z_{2}$ ``charge" of the fracton and dimension-1 excitations (i.e. the presence or absence of an excitation), respectively.   

 \begin{figure}
 $\begin{array}{cc}
 \includegraphics[trim = 0 0 0 0, clip = true, width=0.22\textwidth, angle = 0.]{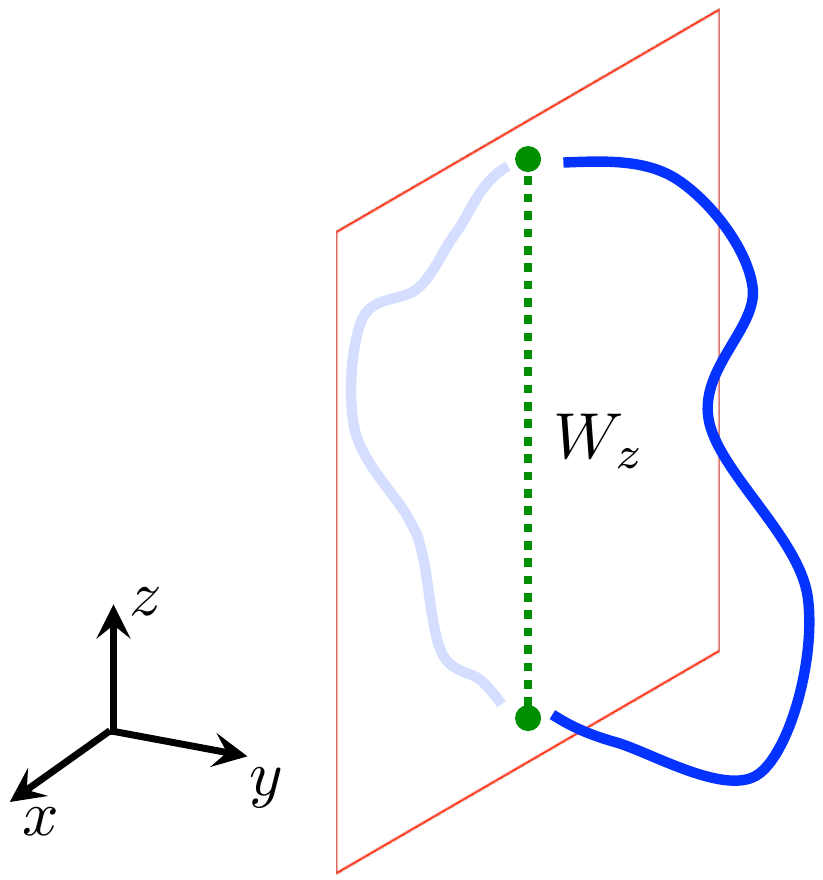}  
 &  \hspace{.1in} \includegraphics[trim = 0 0 0 0, clip = true, width=0.23\textwidth, angle = 0.]{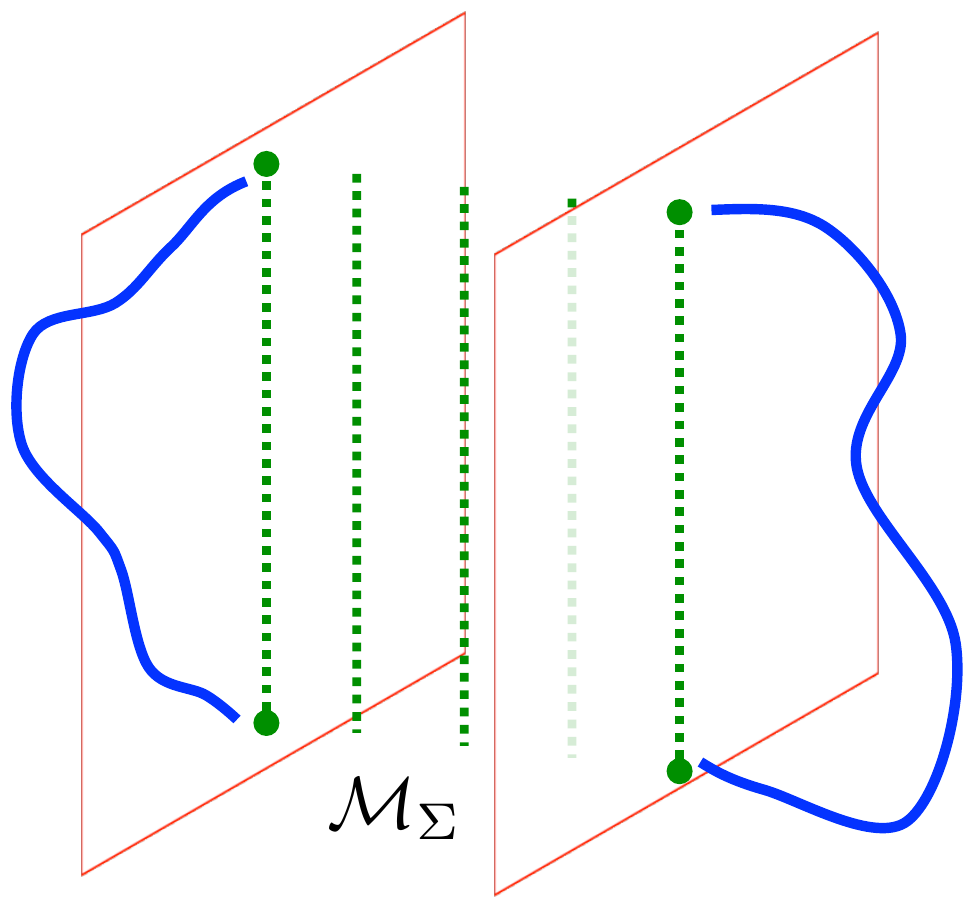} \\
 \text{(a)} & \hspace{.1in}\text{(b)}
 \end{array}$
  \caption{{\bf Loop-Gas Representation}: Acting on the X-cube ground-state  with (a) a straight, line-like operator $W_{z}$ creates a pair of $Z_{2}$ flux excitations at the ends. This operator ``cuts" open a closed loop configuration in the fracton ground-state; alternatively, this operator adds two open strings at its endpoints, when acting on an empty region. 
 For the indicated loop configuration, acting with (b) a membrane operator $\mathcal{M}_{\Sigma}$ yields two separated, open strings of composite flux.  An isolated string endpoint cannot be moved on its own, or else it would be possible to generate a configuration of broken loops that violate the global constraints explained in Sec. IIB.  These isolated excitations are the immobile fractons.  }
  \label{fig:Loop_Gas}
\end{figure}

\subsubsection{Loop Gas Wavefunction, Excitations and Degeneracy}
Our layer construction provides insight into the origin of the exotic topological excitations and sub-extensive ground-state degeneracy of the X-cube phase.  First, the ground-state wavefunction of the X-cube fracton phase may be written as a superposition of composite flux loop excitations of the layered 2D toric codes; heuristically 
\begin{align}
\ket{\Psi_{\mathrm{X-cube}}} \sim \sum_{\mathcal{C}}\ket{\mathcal{C}}
\end{align}
with $\mathcal{C}$, a configuration of loops on the dual lattice. Excitations may be obtained at the end-points of ``broken" composite flux loops.  We note, however, that while any closed loop configuration is permissible in the ground-state, the manner in which these loops may be broken is highly constrained.  First, a composite flux loop that has broken at a single point is precisely a 2D $Z_{2}$ flux excitation of the underlying toric codes. Since these excitations must appear in pairs, the composite flux loops can only be broken at \emph{pairs} of points along $xy$, $yz$ or $xz$ planes.  As an example, acting on the ground-state with $W_{z} = \prod_{\rb,\rb'}\tau^{x}_{\rb\rb'}$, a string operator along a line in the $z$-direction, has the effect of cutting composite flux loops at two points, as shown in Fig. \ref{fig:Loop_Gas}a,
and the resulting pair of 2D $Z_{2}$ fluxes may move within a plane.  This operator creates the mobile anyon excitations in the X-cube phase, which clearly inherit their  ``topological" properties (i.e. self-statistics and mutual statistics with the dimension-1 quasiparticle), from the 2D $Z_{2}$ fluxes in the toric code. 

An isolated string endpoint is precisely the fracton excitation of the X-cube model, and may be obtained by spatially separating the broken endpoints of composite flux loops.  An isolated string endpoint cannot be moved on its own without creating additional excitations, for if such an endpoint were mobile, then it would be possible to smoothly deform a configuration of open strings into one which violates the previously-derived constraint.  The immobile, isolated endpoints then describe the fracton excitations of the X-cube phase.  In practice, acting on the ground-state with a flat membrane operator $\mathcal{M}_{\Sigma} = \prod_{\rb,\rb'\in\Sigma}\tau^{x}_{\rb\rb'}$ will localize these excitations at the corners of $\Sigma$, as shown in Fig. \ref{fig:Loop_Gas}b. In the decoupled toric codes, this operator creates a sequence of 2D $Z_{2}$ fluxes in parallel layers.  
After condensing the composite flux loops, however, only the ends of this layered excitation carry an energy cost, since the bulk of this excitation is locally indistinguishable from a composite flux loop. 

The topological degeneracy of the X-cube fracton topological phase -- which was previously obtained \cite{Fracton_Gauge_Theory} by counting independent constraints on the operators $\mathcal{O}_{\rb}$, $A_{\rb}^{(j)}$ on the three-torus using an algebraic representation of the solvable Hamiltonian \cite{Haah, Polynomial} -- may now be understood by counting the number of topological sectors of the decoupled, two-dimensional toric codes that are made equivalent after condensation of the composite flux loop.  A more extensive discussion of this counting is presented in the Supplemental Material \cite{SM}.

\subsection{$Z_{N}$ X-cube Topological Phase}
The spirit of our proposal motivates the construction of new, three-dimensional topological orders, by appropriately condensing ``loop" objects in an array of two-dimensional topological phases.  An extended discussion of new topological phases that may be built in this manner is the subject of forthcoming work \cite{SV_new}.  Here, we discuss the simplest generalization of our construction, involving an array of two-dimensional $Z_{N}$ toric codes \cite{Kitaev} in the same configuration as shown in Fig. \ref{fig:3D_Stack}, which gives rise to a $Z_{N}$ generalization of the X-cube model.  After arranging the $Z_{N}$ toric codes in a three-dimensional array, each link of the three-dimensional cubic lattice now has two $Z_{N}$ qudit degrees of freedom, denoted $X_{\rb\rb'}$, $Z_{\rb\rb'}$ and  $\widetilde{X}_{\rb\rb'}$, $\widetilde{Z}_{\rb\rb'}$ respectively, and satisfying the algebra $XZ = \omega ZX$, $\widetilde{X}\widetilde{Z} = \omega \widetilde{Z}\widetilde{X}$ with $\omega = e^{2\pi i/N}$. We now consider the Hamiltonian 
\begin{align}\label{eq:ZN_H}
H' = H_{0}' + H_{1}'\
\end{align}
where $H_{0}$ describes the de-coupled layers of $Z_{N}$ toric codes
\begin{align}
H_{0}' = -J\sum_{\rb}\sum_{j=xy,yz,xz}\left[A_{\rb}^{(j)} + B_{\rb}^{(j)}\right] + \mathrm{h.c.}
\end{align}
with $A_{\rb}^{(j)}$ and $B_{\rb}^{(j)}$ the $Z_{N}$ charge and flux operators at site $\rb$, for the toric code oriented along plane $j$, respectively.  
The layers are coupled through the interactions
\begin{align}
H_{1}' = &- h\sum_{\langle \boldsymbol{r},\boldsymbol{r}'\rangle}\left[Z^{\dagger}_{\boldsymbol{r}\boldsymbol{r}'}\widetilde{Z}_{\boldsymbol{r}\boldsymbol{r}'} + \mathrm{h.c.}\right]\\
&- t\sum_{\langle \boldsymbol{r},\boldsymbol{r}'\rangle}\left[X_{\boldsymbol{r}\boldsymbol{r}'}\widetilde{X}_{\boldsymbol{r}\boldsymbol{r}'} + \mathrm{h.c.}\right] \nonumber
\end{align}

 \begin{figure}
 $\begin{array}{cc}
 \includegraphics[trim = 0 0 0 0, clip = true, width=0.14\textwidth, angle = 0.]{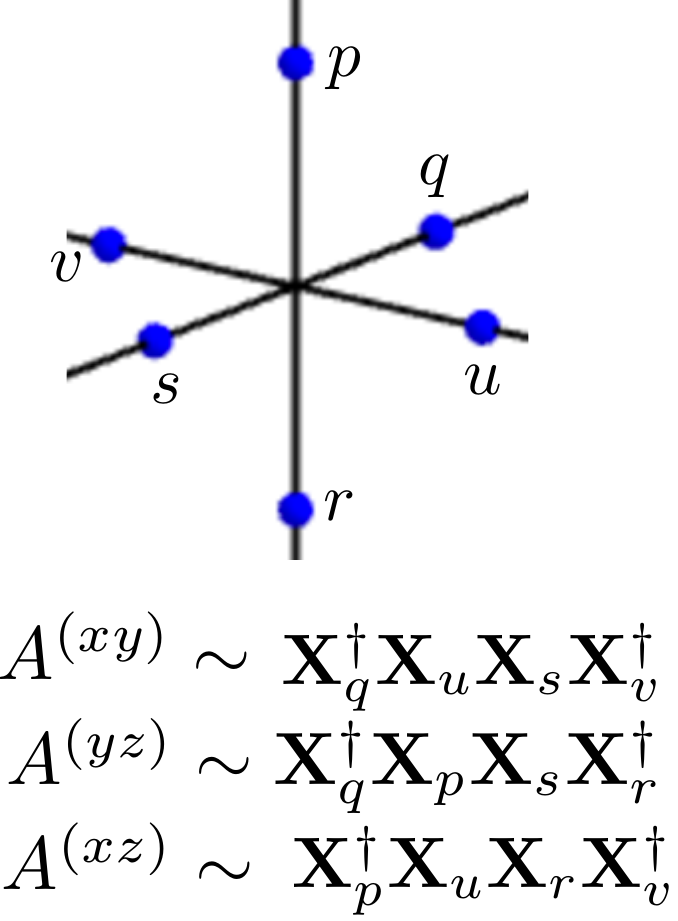}  
 &  \includegraphics[trim = 0 0 0 0, clip = true, width=0.27\textwidth, angle = 0.]{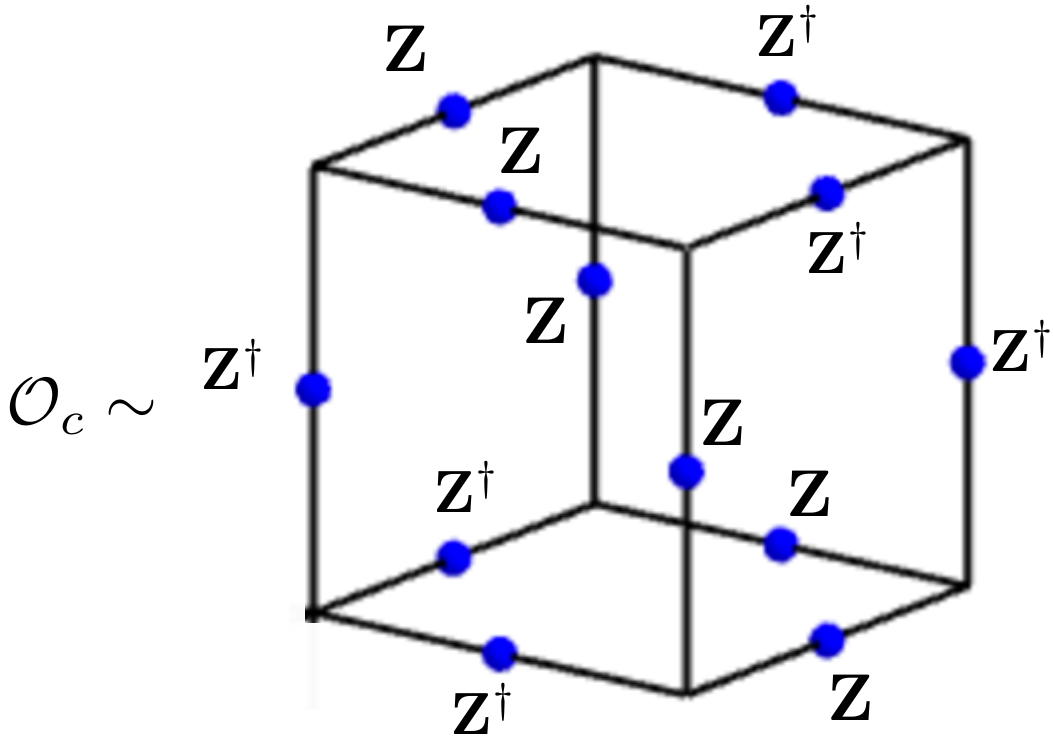} 
 \end{array}$
  \caption{{\bf The $Z_{N}$ X-cube Model}: Shown are the commuting operators that appear in the solvable Hamiltonian for the $Z_{N}$ X-cube model $H = -K\sum_{c}(\mathcal{O}_{c} + \mathcal{O}_{c}^{\dagger}) - J\sum_{\rb,j}(A_{\rb}^{(j)} + A_{\rb}^{(j)\dagger})$.  The $Z_{2}$ case discussed in Sec. IB, is obtained by replacing $\boldsymbol{Z}$, $\boldsymbol{Z}^{\dagger}\rightarrow\tau^{z}$ and $\boldsymbol{X}$, $\boldsymbol{X}^{\dagger}\rightarrow\tau^{x}$. }
  \label{fig:Z_N_Xcube}
\end{figure}

 \begin{figure*}
  $\begin{array}{ccc}
 \includegraphics[trim = 0 -30 0 0, clip = true, width=0.43\textwidth, angle = 0.]{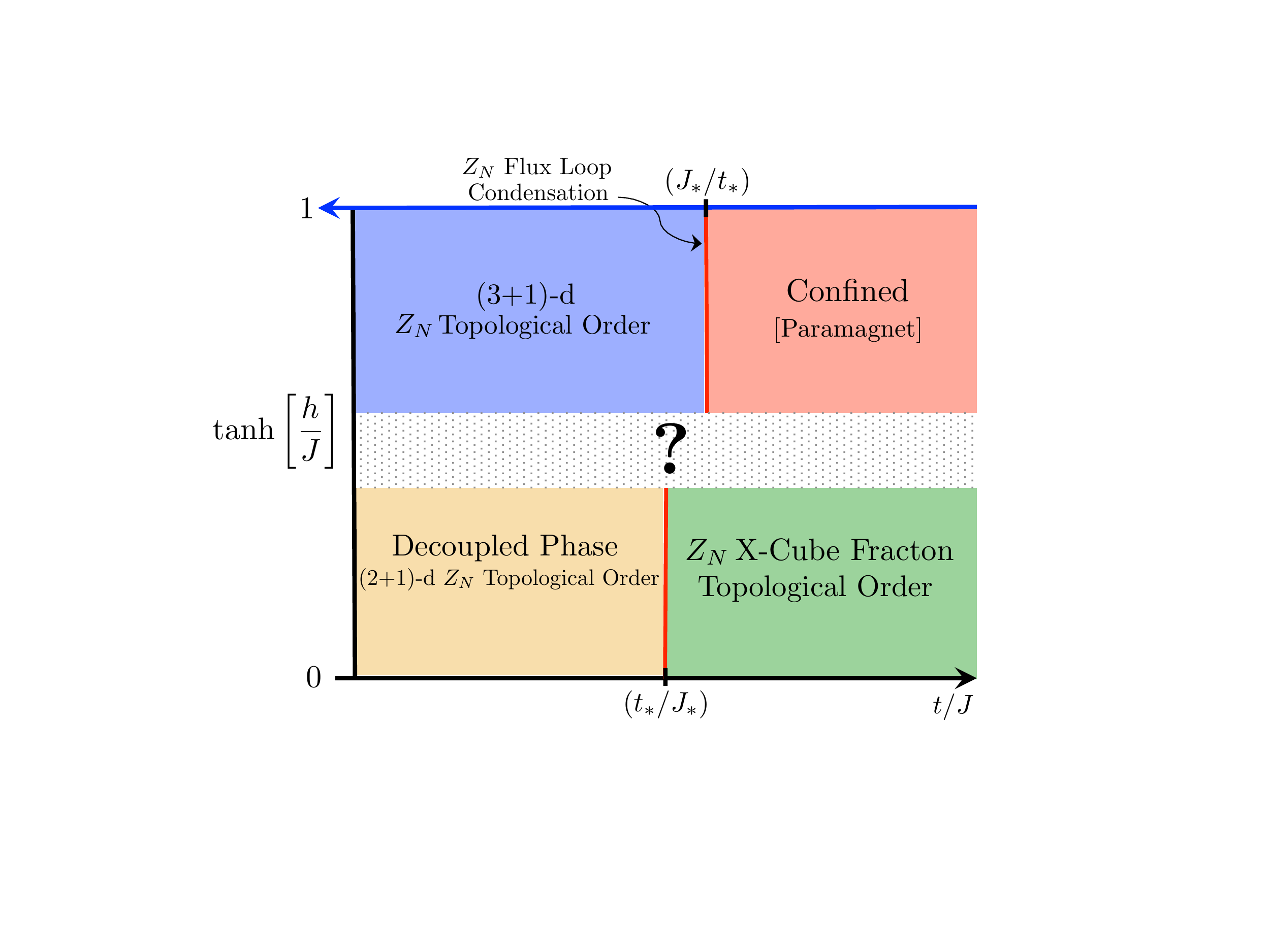} & &
 \hspace{.5in} \includegraphics[trim = 0 -10 0 0, clip = true, width=0.43\textwidth, angle = 0.]{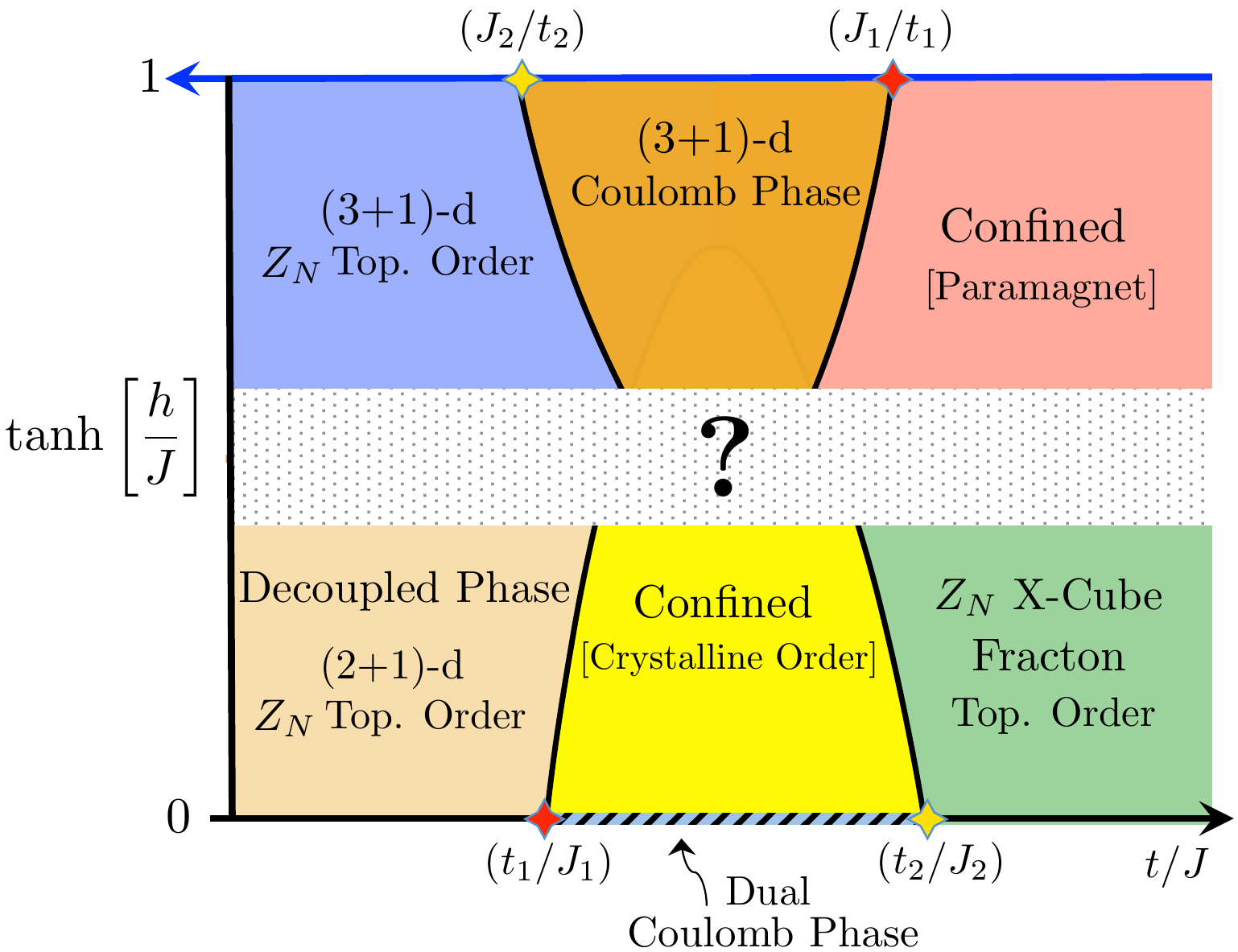} \\
 \text{\hspace{.4in}(a)\hspace{.1in}} \boldsymbol{N < 5} & & \hspace{.5in}
  \text{\hspace{.4in}(b)\hspace{.1in}} \boldsymbol{N \ge 5}
  \end{array}$
 \caption{{\bf Schematic Phase Diagram of the Coupled System}:  
The duality derived in Sec. II relates the points $(x,y) = (t/J,0)$ with $(J/t, 1)$ that lie along the black and blue arrows, respectively, in the schematic phase diagrams above. From knowledge of the phase diagram of (3+1)-d  $Z_{N}$ lattice gauge theory, we argue that when $N < 5$, there is a direct, first-order transition between the X-cube phase and the decoupled phase.  Alternatively, when $N \ge 5$, there must be an intermediate phase along the line $h = 0$ which is dual to a Coulomb phase that appears in the phase diagram of the (3+1)-d $Z_{N}$ lattice gauge system. We argue that this phase is unstable \cite{Xu_Wu} and becomes a gapped, topologically trivial phase when $h > 0$.  From Monte Carlo studies of the transition along the blue arrow, between the deconfined phase of $Z_{N}$ gauge theory and the Coulomb phase \cite{Rebbi_Numerics_1}, we believe that the dual transition along the $h = 0$ line, into the fracton topological phase in (b), is continuous. }
  \label{fig:Phase_Diagram}
\end{figure*}

Following our previous analysis, we consider the limit $h \gg J$, $t$, where we find that the ground-state exhibits 3D $Z_{N}$ topological order.  In this limit, we may identify a single $Z_{N}$ degree of freedom on each link of the cubic lattice (with logical operators $\boldsymbol{X}$, $\boldsymbol{Z}$), and obtain the effective Hamiltonian $H_{\mathrm{toric}}' - t\sum_{\rb\rb'}[\boldsymbol{X}_{\rb\rb'} + \mathrm{h.c.}]$ where $H_{\mathrm{toric}}'$ is now the 3D $Z_{N}$ toric code Hamiltonian \cite{3D_Toric}.  Alternatively, when $t \gg J$, $h$ so that we have condensed the ``$Z_{N}$ composite flux loop", we obtain the effective Hamiltonian
\begin{align}
H_{\mathrm{eff}}' = H_{\mathrm{X}\text{-}\mathrm{cube}_{N}} - h\sum_{\langle\rb,\,\rb'\rangle}\left[\boldsymbol{Z}_{\rb\rb'} + \boldsymbol{Z}_{\rb\rb'}^{\dagger}\right]
\end{align}
where
\begin{align}\label{eq:H_ZN}
H_{\mathrm{X}\text{-}\mathrm{cube}_{N}} = -K\sum_{c}[\mathcal{O}_{c} + \mathcal{O}_{c}^{\dagger}] - J\sum_{\rb,\,j}[A_{\rb}^{(j)} + (A_{\rb}^{(j)})^{\dagger}]
\end{align}
is the solvable Hamiltonian for the $Z_{N}$ generalization of the X-cube model.  The microscopic form of the operators $\mathcal{O}_{c}$ and $A_{\rb}^{(j)}$ are explicitly given in Fig. \ref{fig:Z_N_Xcube}.  

Since the operators $\mathcal{O}_{c}$ and $A_{\boldsymbol{r}}^{(j)}$ commute, the ground-state of (\ref{eq:H_ZN}) again satisfies $\mathcal{O}_{c}\ket{\Psi} = A_{\boldsymbol{r}}^{(j)}\ket{\Psi} = \ket{\Psi}$, and gapped excited states may be created by acting with straight Wilson line or flat membrane operators, in a straightforward generalization of the operators in the $Z_{2}$ X-cube model.  Now however, the dimension-1 quasiparticles and fracton excitations carry a $Z_{N}$ electric and magnetic charge, respectively, which is inherited from the underlying 2D $Z_{N}$ toric codes.  As a result, a pair of fractons which are mobile in a plane have bosonic self-statistics, but non-trivial mutual statistics $e^{i\theta_{pq}}$ with a dimension-1 quasiparticle of charge $q \in \{1,\ldots, N-1\}$ in its plane of motion. Here, the statistical angle $\theta_{pq} = 2\pi pq/N$, where $p$ is the magnetic charge carried by the fracton excitation. 

%
%
%

\section{Phase Diagram}

We now turn to a discussion of the rich phase diagram of the coupled $Z_{N}$ toric codes, as described by the Hamiltonian (\ref{eq:ZN_H}).  Our results are summarized in the diagrams shown in Fig. \ref{fig:Phase_Diagram}a and b when $N < 5$ and $N\geq 5$, respectively.  Both figures include the four distinct phases discussed previously, i.e.
\begin{enumerate}[label=]
\item {\bf I.} \underline{$J \gg h$, $t$:} \,\,\,Decoupled 2D $Z_{N}$ Topological Phase
\item {\bf II.} \underline{$h\gg J$, $t$:} \,\,\,3D $Z_{N}$ Topological Phase
\item {\bf III.} \underline{$t \gg J$, $h$:} \,\,\,$Z_{N}$ X-Cube Fracton Phase
\item {\bf IV.} \underline{$h$, $t\gg J$:} \,\,\,Trivial, Confined Phase 
\end{enumerate} 
The remaining details in the phase diagram arise from an interesting duality between the Hamiltonian (\ref{eq:ZN_H}) when $h = 0$, and the 3D $Z_{N}$ toric code in the presence of a field; increasing the strength of this field eventually leads to condensation of the $Z_{N}$ flux loops and results in a trivial, confined phase.  From this duality, and from knowledge of the phase structure of $Z_{N}$ lattice gauge theory, we deduce that when $N < 5$, the $Z_{N}$ X-cube model has a direct, first-order transition to a phase where the layers are effectively decoupled.  When $N \ge 5$, however, there is an intermediate phase which is dual to a Coulomb phase that is known to appear in the phase diagram of (3+1)-d $Z_{N}$ lattice gauge theory \cite{Elitzur,  Guth}, along the line $h = 0$.  We argue, however, that this phase is gapped and topologically trivial when $h > 0$.  Interestingly, one of the transitions out of the gapless phase is believed to be {continuous} in numerical studies \cite{Rebbi_Numerics_1, Rebbi_Numerics_2}, which presents the intriguing possibility that there exists a continuum field theory description for the $Z_{N}$ X-cube phase.  

Several features of the phase diagrams in Fig. \ref{fig:Phase_Diagram}, including 
the manner in which these phases meet at the center of the phase diagram, as well as the generic behavior of the transitions between the $Z_{N}$ X-cube and trivial confined phases, are unknown. We conclude by presenting a solvable projector model in which the transition is believed to be first-order, and by speculating on directions for future work.

\subsection{Confinement Transition from the $Z_{N}$ Topological Phase}
We begin from the top left corner of the phase diagram ($h \gg J$, $t$), where our system is in a 3D $Z_{N}$ topological phase.  Increasing the strength of the coupling $t$ eventually leads to a condensation of $Z_{N}$ flux loops, and the associated phase transition is captured by the finite-temperature behavior of the 4D classical $Z_{N}$ lattice gauge theory with a ``Wilsonian" imaginary time action $S_{W}[\theta] = \beta\widetilde{K}\sum_{p}\cos[(\Delta\times\theta)_{p}]$, where $\theta$ is a $Z_{N}$ variable defined on the links of the 4D cubic lattice, while $(\Delta\times\theta)_{p}$ is the lattice curl around plaquette $p$, and the sum is over all plaquettes on the 4D lattice.  

When $N < 5$, classical Monte Carlo studies \cite{Rebbi_Numerics_1, Rebbi_Numerics_2} have observed a direct, first-order phase transition between a disordered phase where the magnetic flux loops have condensed at high temperatures, to one where they remain energetically costly, with the expectation value of classical Wilson loop operators decaying as the exponential of the area of an enclosed region (``area-law") and as their total length (``perimeter-law"), respectively, in the two phases. In our problem, these two phases correspond to a trivial loop condensate and a phase with $Z_{N}$ topological order, respectively, as indicated in the top portion of the phase diagram in Fig. \ref{fig:Phase_Diagram}a.

When $N$ is large ($N \ge 5$), \emph{three} phases are observed in Monte Carlo studies of the classical 4D $Z_{N}$ lattice gauge theory \cite{Rebbi_Numerics_1, Rebbi_Numerics_2, Svetitsky}. This rich phase structure may be understood in a variety of ways.  To begin, we recall that a similar phenomenon occurs in 2D classical, $Z_{k}$-symmetric spin models when $k$ is sufficiently large. In addition to a low-temperature symmetry-breaking phase and a disordered phase at high temperatures, there is an intermediate  phase with algebraic correlations that resembles the low-temperature behavior of the XY model.  The existence of this XY-like phase with an emergent global $U(1)$ symmetry may be motivated by recalling that in the classical 2D XY model, $Z_{k}$ anisotropy is a  marginally irrelevant perturbation when $k \ge 4$ \cite{Kadanoff}.  A more rigorous argument for the existence of this phase, even when the anisotropy is infinitely strong, is provided in Ref. \cite{Elitzur}.

A similar argument may be used for the presence of an intermediate phase with emergent $U(1)$ gauge symmetry in the phase diagram of 4D $Z_{N}$ lattice gauge theory in the large-$N$ limit \cite{Svetitsky, Elitzur, Guth}. Ref. \cite{Guth} studied a Villain form of the 4D $Z_{N}$ gauge theory, which exhibits a discrete analog of the  electric-magnetic self-duality of compact $U(1)$ gauge theory \cite{Casher}, and whose action may be re-written as that of Maxwell electrodynamics in the presence of both electric and magnetic charges.  While the action exhibits only a $Z_{N}$ gauge symmetry, it was argued that  both the electric and magnetic excitations are strongly suppressed near the self-dual point in the large-$N$ limit, so that the effective action is that of pure Maxwell electromagnetism.   As a result, when $N \ge 5$, there is a ``Coulomb phase" that is encountered in the top region of the phase diagram in Fig. \ref{fig:Phase_Diagram}b, with the $Z_{N}$ variable $\theta$ effectively behaving as the emergent $U(1)$ gauge field.

These three phases may be distinguished by the behavior of the gauge-invariant Wilson loop operator $W_{C}$, which creates a closed tube of electric flux, along with its (electric-magnetic) dual $\Gamma_{\widetilde{C}}$, which is a membrane-like operator that creates a closed magnetic flux loop along loop $\widetilde{C}$ on the dual lattice.  In the deconfined phase of the $Z_{N}$ gauge theory, a large Wilson loop $W_{C}$ exhibits perimeter-law behavior while $\Gamma_{\widetilde{C}}$ has area-law behavior, while the opposite behavior occurs in the confined phase.  In the intermediate phase, however, both charge and loop excitations are suppressed \cite{Guth} and neither excitation has condensed.  As a result, the Wilson loop operator $W_{C}$ must exhibit perimeter-law behavior \cite{Guth}. Since this phase is self-dual under the electric-magnetic duality transformation, the \emph{same} behavior holds for $\Gamma_{\widetilde{C}}$.  As argued in Ref. \cite{Hooft}, due to the non-trivial statistics of $Z_{N}$ charges and flux loops, perimeter-law behavior for both operators can only occur in the presence of a gapless gauge field, which is consistent with the emergence of a Coulomb phase. 
  
It remains unclear whether the transition between the Coulomb and confined phases is continuous \cite{Jersak_1, Jersak_2} or weakly first-order \cite{QED, QED_1}.  However, the transition between the deconfined phase of the $Z_{N}$ gauge theory and the Coulomb phase appears continuous in Monte Carlo studies \cite{Rebbi_Numerics_1, Rebbi_Numerics_2, Svetitsky}.  Finally, as $N$ increases, the region of stability for the $Z_{N}$ topological phase shrinks, while the second transition from the Coulomb phase to the confined phase remains robust \cite{Rebbi_Numerics_1}.  This is consistent with the expectation that the limit ``$N\rightarrow +\infty$", should somehow reproduce the phase diagram of the pure (3+1)-d compact $U(1)$ gauge theory.  

\subsection{Transition(s) from the decoupled phase to the $Z_{N}$ X-cube phase}


We now study the bottom half of the phase diagrams shown in Fig. \ref{fig:Phase_Diagram}a \& b.  We first demonstrate that the transition from the decoupled layers of 2D toric codes to the X-cube fracton phase is \emph{dual} to the confinement transition for the (3+1)-d $Z_{N}$ gauge theory, which is driven by the condensation of flux loops.  This result is intuitively apparent from our ``loop-gas" representation of the X-cube phase, which suggests that the transition from the decoupled layers must be driven by the condensation of a loop excitation.  Physically, our duality transformation maps the closed composite flux loops in the ground-state of the X-cube phase into the closed electric flux loops in the ground-state of the $Z_{N}$ topological phase.  We emphasize that this duality relates the bulk spectrum of two seemingly distinct physical systems, and provides relations between certain local operators on both sides.  

We begin by explicitly implementing our lattice duality transformation on the Hamiltonian for the coupled $Z_{2}$ toric codes (\ref{eq:H}) when $h = 0$, before discussing its consequences.  In practice, our transformation is identical to the (2+1)-d Wegner duality that maps the transverse-field Ising model to the (2+1)-d $Z_{2}$ gauge theory \cite{Wegner}, in \emph{each layer} of our coupled system; as a result, the generalization of this duality for the $Z_{N}$ case is apparent from our following discussion.  Recall that the Hamiltonian (\ref{eq:H}) for the coupled toric codes when $h = 0$ is given by
\begin{align}\label{eq:H_h_0}
H = -J\sum_{\rb,\,j}\left[A_{\rb}^{(j)} + B_{\rb}^{(j)}\right] - t\sum_{\langle \rb, \rb'\rangle}\sigma^{x}_{\rb\rb'}\mu^{x}_{\rb\rb'}
\end{align}
We recognize that the 2D $Z_{2}$ charge operator $A_{\rb}^{(j)}$ and the operator $\mathcal{O}_{\rb}=B_{\rb}^{(xy)}B_{\rb}^{(yz)}B_{\rb}^{(xz)}B_{\rb + \hat{x}}^{(yz)}B_{\rb + \hat{y}}^{(xz)}B_{\rb + \hat{z}}^{(xy)}$, which is the product of six flux operators around an elementary cube, as in Eq. (\ref{eq:Cube_Op}), commute with the Hamiltonian (\ref{eq:H_h_0}) and with each other $[\mathcal{O}_{\rb}, A_{\rb'}^{(j)}] = [A_{\rb}^{(j)}, H] = [\mathcal{O}_{\rb}, H] = 0$. Therefore, we work in a restricted Hilbert space where
\begin{align}
A_{\rb}^{(j)}\ket{\Psi} = \ket{\Psi}\hspace{.2in} \text{and} \hspace{.2in} \mathcal{O}_{\rb}\ket{\Psi} = \ket{\Psi}
\end{align}
without loss of generality.

\begin{figure}
 \includegraphics[trim = 0 0 0 0, clip = true, width=0.47\textwidth, angle = 0.]{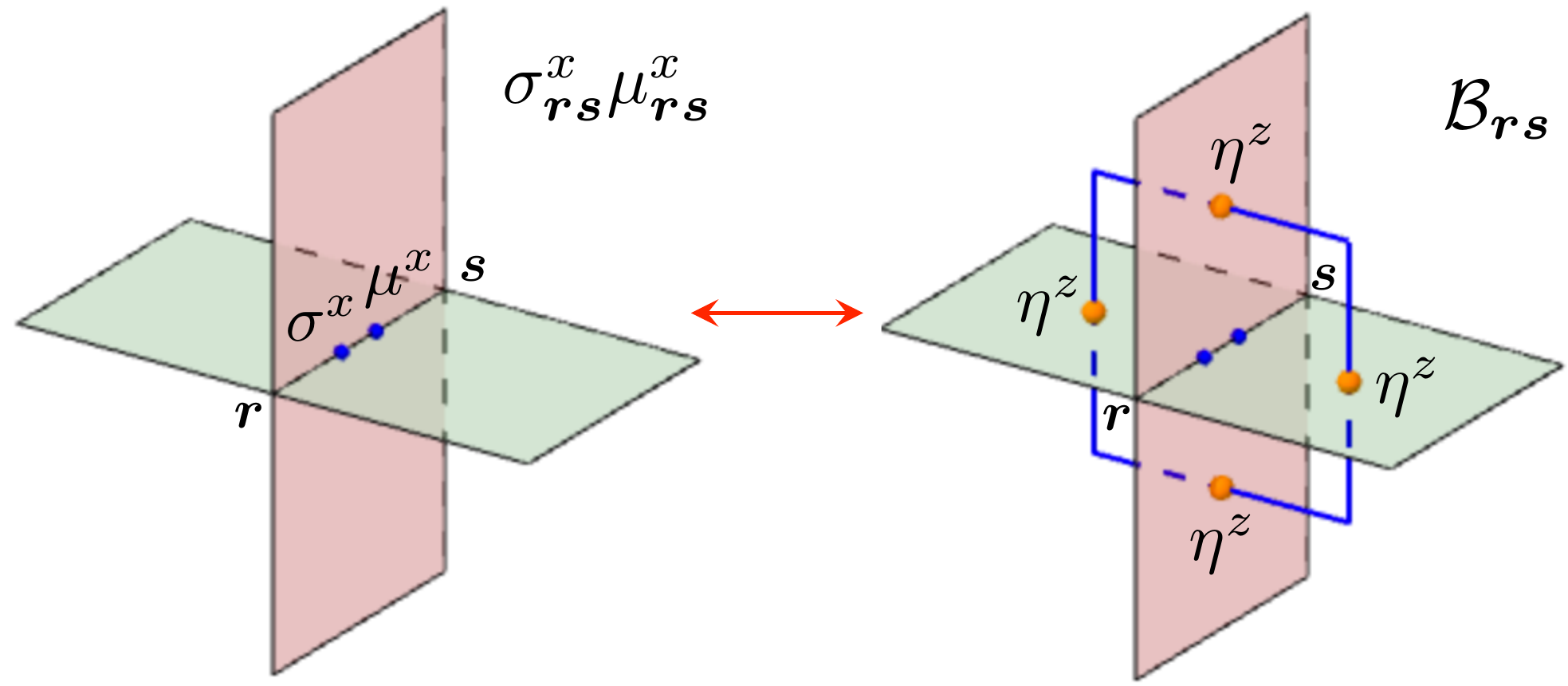}  
 \caption{{\bf Bond-Plaquette Duality}: We provide a dual representation of the Hamiltonian (\ref{eq:H}) by introducing spins ($\eta$) on the links of the dual cubic lattice, which measure the flux through an elementary plaquette in the decoupled layers.  This dual representation allows us to demonstrate that the transition from the decoupled layers to the $Z_{N}$ X-cube phase is dual to a confinement transition in $Z_{N}$ gauge theory.}
  \label{fig:Duality}
\end{figure}

We now introduce a {dual} description of the transition from the decoupled theory ($t \ll J$) to the X-cube phase ($J \gg t$) that solves the first of these constraints, by introducing spins $(\eta)$ on the links of the dual cubic lattice. These spins are to be interpreted as measuring the flux through a plaquette in the decoupled toric code layers.  We implement the dual representation by performing the replacements
\begin{align}
B_{\rb}^{(j)} \,&\longrightarrow\,\eta^{x}_{\rb,j}\\
\sigma^{x}_{\rb\sB}\mu^{x}_{\rb\sB}\,&\longrightarrow\,\mathcal{B}_{\rb\sB} \equiv \prod_{\rb',j\in\mathrm{plaq}_{\rb\sB}}\eta^{z}_{\rb',j}
\end{align}
where $\mathcal{B}_{\rb\sB}$ is a four-spin operator on a plaquette of the dual lattice  pierced by the link $\langle \rb,\,\sB\rangle$ as shown in Fig. \ref{fig:Duality}.     

The duality transformation preserves the algebra of local operators in (\ref{eq:H_h_0}), and naturally solves the constraint $A_{\rb}^{(j)}\ket{\Psi} = \ket{\Psi}$.  Furthermore, the dual description of the operator $\mathcal{O}_{\rb}$ is given by
\begin{align}\label{eq:Gauss}
\mathcal{O}_{\rb} \longrightarrow \mathcal{A}_{\rb} = \prod_{\sB,j\in\mathrm{star}_{\rb}}\eta^{x}_{\sB,j}
\end{align}
where $\mathcal{A}_{\rb}$ is precisely the six-spin $Z_{2}$ charge operator which acts along a ``star" configuration of the $\eta$ spins at each site of the dual cubic lattice. We therefore find that the dual Hamiltonian is precisely
\begin{align}\label{eq:H_dual}
H_{\mathrm{dual}} = -J\sum_{\rb, j}\eta^{x}_{\rb, j} - t\sum_{\langle\rb, \rb'\rangle}\mathcal{B}_{\rb\rb'}
\end{align}
supplemented by the $Z_{2}$ Gauss's law constraint on the dual cubic lattice
\begin{align}
\mathcal{A}_{\rb}\ket{\Psi} = \ket{\Psi}
\end{align}
which describes the confinement transition for a 3D $Z_{2}$ topological phase driven by the condensation of flux loops.  As advertised, the lattice duality relates the $Z_{2}$ X-cube phase to the deconfined phase of (3+1)-d $Z_{2}$ gauge theory, while the decoupled toric code layers are dual to the trivial, confined phase.  

Our duality transformation establishes that the bulk transition from the decoupled toric codes to the X-cube fracton phase is dual to flux loop condensation in the 3D $Z_{2}$ gauge theory, which is known to be a direct, first-order transition.  The natural $Z_{N}$ generalization of this duality transformation leads us to conclude that along the $h = 0$ line of the phase diagram (i) when $N < 5$, there is a direct, first-order transition between the decoupled layers and the X-cube phase and (ii)  when $N \ge 5$, there is an intermediate gapless phase for the layered system, which is dual to the Coulomb phase that emerges in the phase diagram of the (3+1)-d $Z_{N}$ gauge theory.  
From our duality transformation, it appears that this gapless ``dual Coulomb" phase along the $h = 0$ line of the phase diagram is characterized by an emergent $U(1)$ gauge symmetry, which should be generated by the natural generalization of the Gauss's law condition $\mathcal{O}_{\rb} = +1$ for the X-cube fracton phase.  The $U(1)$ generalization of this condition may be read off from Fig. \ref{fig:Z_N_Xcube}; if we consider an integer-valued ``electric field" tensor $E_{ij}$ with vanishing diagonal components $E_{ii} = 0$, and a conjugate gauge field $A_{ij}$, then this Gauss's law takes the simple form 
\begin{align}
\Delta_{i}\Delta_{j}E_{ij} = 0
\end{align} 
with $\Delta_{i}$, the lattice derivative along direction $i$, and summation performed over repeated indices.  

The ``higher-rank" compact $U(1)$ gauge theory \cite{Pretko, Xu} defined by this Gauss's law condition is equivalent to a generalized dimer model studied in Ref. \cite{Xu_Wu}, which was shown to be in a gapped phase with crystalline order due to the proliferation of topological defects in the gauge field configurations (i.e. ``monopole" events).  Due to this result, we believe that the dual Coulomb phase is unstable, and gives way to a gapped, topologically trivial phase with conventional long-range order when $h > 0$.  The $h = 0$ line then corresponds to the fine-tuned limit where monopole events in the gauge field are suppressed.  This conclusion is consistent with the fact that the perturbation $\sigma^{z}_{\rb\rb'}\mu^{z}_{\rb\rb'}$ corresponds to a highly non-local operator in terms of the dual $\eta$ spins, and with our expectation for the behavior of the phase diagram in the $N\rightarrow +\infty$ limit.  
The nature of the transitions between this trivial phase, the X-cube phase and the decoupled layers are unknown when $h > 0$.  When $h = 0$, however, our duality mapping suggests that the phase transition from the X-cube phase to the dual Coulomb phase is dual to the transition between the deconfined phase of (3+1)-d $Z_{N}$ gauge theory and the ordinary Coulomb phase, which may be continuous \cite{Svetitsky, Rebbi_Numerics_1}, though we are unaware of the continuum field theory that governs the properties of this transition. 

We conclude this section by identifying the following order parameters 
\begin{align}
{W}_{\Sigma} \equiv \prod_{\langle\rb,\,\sB\rangle\in\Sigma} X_{\rb\sB}\widetilde{X}_{\rb\sB}\hspace{.4in}
{\Gamma}_{\Sigma} \equiv \prod_{(p,j)\in\Sigma} B_{p}^{(j)}
\end{align}
that distinguish the X-cube phase, the decoupled phase and the dual Coulomb phase along the line $h = 0$. Here, $\Sigma$ is a two-dimensional region with area $A_{\Sigma}$ and perimeter $P_{\Sigma}$. The product appearing in the definition of $W_{\Sigma}$ is taken along the bonds perpendicular to the region $\Sigma$, and this operator may be thought of as inserting a composite flux loop along the boundary of $\Sigma$. Furthermore, ${\Gamma}_{\Sigma}$ is the product of the plaquette operators that measure the 2D $Z_{N}$ flux through the region $\Sigma$.  Our notation is meant to emphasize that these operators are dual to the Wilson loop and membrane operators that distinguish the various phases of the $Z_{N}$ gauge theory. From our discussion in Sec. IIA, we determine that the operators $W_{\Sigma}$, $\Gamma_{\Sigma}$ exhibit the following behavior along the $h = 0$ line, when the region $\Sigma$ is sufficiently large:
\begin{itemize}
\item Decoupled Phase: $\langle{W}_{\Sigma}\rangle \sim e^{-\sigma A_{\Sigma}}$,\,\,\, $\langle{\Gamma}_{\Sigma}\rangle \sim e^{-c P_{\Sigma}}$\\
\item Dual Coulomb Phase: $\langle{W}_{\Sigma}\rangle$, $\langle{\Gamma}_{\Sigma}\rangle \sim e^{-c P_{\Sigma}}$\\
\item X-cube Phase: $\langle{W}_{\Sigma}\rangle \sim e^{-c P_{\Sigma}}$,\, \,\,$\langle{\Gamma}_{\Sigma}\rangle \sim e^{-\sigma A_{\Sigma}}$ 
\end{itemize}


\subsection{Confinement of the X-cube fracton phase}

The generic behavior of the confinement transition for the $Z_{N}$ X-cube fracton topological phase -- by condensing the dimension-1 quasiparticle excitations -- are not known.  This transition occurs in the far right region of the phase diagrams in Fig. \ref{fig:Phase_Diagram}.  Instead of studying this transition directly, however, we propose a solvable Hamiltonian that interpolates between the $Z_{2}$ X-cube fracton topological order and a confined phase in this section and determine that the transition is first-order within this model. It is unknown whether this captures the generic behavior of this transition, outside of the solvable model.

We construct a solvable projector model by placing spins $(\tau)$ on the links of a three-dimensional cubic lattice.  Now, however, we consider the Hamiltonian
\begin{align}\label{eq:H_Solvable}
\mathcal{H} = -J\sum_{\rb, j}A_{\rb}^{(j)} + J\sum_{\rb, j}\left[ \prod_{\sB\in\mathrm{plane}_{j}(\rb)}e^{-h\tau^{z}_{\rb\sB}/J}\right]
\end{align}
supplemented by the constraint
\begin{align}
\mathcal{O}_{\rb}\ket{\Psi} = \ket{\Psi}.
\end{align}
where $A_{\rb}^{(j)} = \prod_{\sB\in\mathrm{plane}_{j}(\rb)}\tau^{x}_{\rb\sB}$ is the four-spin operator in the X-cube model that is oriented in the $j$th plane, and measures the presence of a dimension-1 quasiparticle excitation at site $\rb$.  As before, the operator $\mathcal{O}_{\rb}$ is the 12-spin  $\tau^{z}$ operator as defined in Eq. (\ref{eq:Cube_Op}) that measures the fracton ``charge" at a cube
$\mathcal{O}_{\rb} = \prod_{\sB,\,\rb'\in\mathrm{cube}(\rb)}\tau^{z}_{\rb'\sB}$.
We observe that the Hamiltonian $\mathcal{H}$ exhibits both the $Z_{2}$ X-cube fracton phase ($h \ll J$) as well as a trivial confined phase ($h \gg J$).  Furthermore, when $h \ll J$, the Hamiltonian $\mathcal{H}$ reduces to the effective Hamiltonian considered previously for the confinement transition of the X-cube phase
\begin{align}
\mathcal{H}_{\mathrm{eff}} =  -J\sum_{\rb, j}A_{\rb}^{(j)} - 2h\sum_{\langle\rb,\,\rb'\rangle}\tau^{z}_{\rb\rb'} + O(h^{2}/J)
\end{align}

The full Hamiltonian $\mathcal{H}$ (\ref{eq:H_Solvable}) is a sum of projection operators $\Pi_{\rb, j} \equiv -A_{\rb}^{(j)} + \prod_{\sB \in\mathrm{plane}_{j}(\rb)}e^{-h\tau^{x}_{\rb\sB}/J}$ and therefore has a positive semi-definite spectrum.   As a result, a ground-state of the Hamiltonian is found by explicitly constructing a zero-energy wavefunction.  
Projector Hamiltonians similar to (\ref{eq:H_Solvable}) have been extensively studied by considering generalizations of the Rokhsar-Kivelson (RK) point in the two-dimensional quantum dimer model \cite{RK, Fradkin, Henley, Castelnovo_Chamon, Claudio12}, where the ground-state may be written as an equal-amplitude superposition of dimer configurations.  For these generalized RK models, the Hamiltonian is related to the Markovian transition matrix for a classical system which satisfies detailed balance \cite{Henley}.  Spatial correlation functions of certain ``diagonal" operators in the ground-state of the quantum model are identical to classical correlation functions in equilibrium. 

Let $\ket{\psi_{\mathrm{fracton}}}$ be a ground-state of the Hamiltonian $\mathcal{H}$ when $h = 0$, i.e. one of the ground-states of the solvable X-cube fracton Hamiltonian.  We may write $\ket{\psi_{\mathrm{fracton}}}$ explicitly as
\begin{align}
\ket{\psi_{\mathrm{fracton}}} &= \prod_{\rb, j}\left(\frac{1 + A_{\rb}^{(j)}}{2}\right)\ket{\tau^{z}_{\sB\sB'} = +1}
\end{align}
where $\ket{\tau^{z}_{\sB\sB'} = +1}$ is the state with all spins polarized in the $z$-direction.  The ground-state of $\mathcal{H}$ is then given by
\begin{align}
\ket{\Psi_{\mathrm{gs}}} \sim \prod_{\langle \rb,\,\sB \rangle}e^{h\tau^{z}_{\rb\sB}/2J}\ket{\psi_{\mathrm{fracton}}}
\end{align}
It is straightforward to check that $\ket{\Psi_{\mathrm{gs}}}$ is indeed a zero-energy eigenstate of the Hamiltonian, as it is annihilated by all of the projectors $\Pi_{\rb, j}$. 
 
It is convenient to re-cast the ground-state wavefunction in an alternate form.  Observe that configuration of spins appearing in the state $\ket{\psi_{\mathrm{fracton}}}$ may equivalently be specified by the action of the $A_{\rb}^{(j)}$ operators on the reference state $\ket{\tau^{z}_{\sB\sB'} = +1}$.   As a result, the ground-state wavefunction $\ket{\Psi_{\mathrm{gs}}}$ may be re-written in terms of dual Ising variables $\mu$, $\sigma$ at the sites of a three-dimensional cubic lattice, which label the presence or absence of a particular operator $A_{\rb}^{(j)}$ acting on the reference configuration.  In this dual representation, the ground-state takes the form
\begin{align}
\ket{\Psi_{\mathrm{gs}}} = \frac{1}{\sqrt{\mathcal{Z}}}\sum_{\{\sigma\}}\sum_{\{\mu\}}e^{-\beta H/2}\ket{\{\sigma\}, \{\mu\}}
\end{align}
where the classical Hamiltonian $\beta H$ is given by
\begin{align}\label{eq:H_class}
\beta H = \frac{h}{J}\sum_{\rb}\left[\sigma_{\rb}\sigma_{\rb + \hat{x}} + \mu_{\rb}\mu_{\rb + \hat{y}} + \sigma_{\rb}\mu_{\rb}\sigma_{\rb + \hat{z}}\mu_{\rb + \hat{z}}\right]
\end{align}
and $\mathcal{Z}$ is the partition function $\mathcal{Z} = \sum_{\{\sigma\}}\sum_{\{\mu\}} e^{-\beta H}$ for this classical model.

Preliminary Monte Carlo studies of the classical Hamiltonian $\beta H$ reveal a first-order phase transition when $h/J \approx 1.13$ \cite{Johnston_1, Johnston_2}, which implies the presence of a first-order phase transition in our model.  For example, the expectation value of the magnetization $M = \sum_{\langle \rb,\rb'\rangle}\tau^{z}_{\rb\rb'}$ is precisely the energy of the classical system $E = \langle\Psi\,|\,M\,|\,\Psi\rangle$, which exhibits a discontinuity at the putative phase transition point \cite{Johnston_1, Johnston_2}.  Classical correlation functions that are not invariant under the $Z_{2}$ transformation $\sigma \rightarrow -\sigma$, $\mu \rightarrow - \mu$ along any plane of the cubic lattice must vanish, as this is a symmetry of the classical Hamiltonian (\ref{eq:H_class}).  To our knowledge, however, the behavior of higher-point correlation functions near the classical phase transition have not yet been studied.  An intriguing possibility is that the  \emph{dynamical} behavior of this classical system is ``glassy" in a certain range of couplings $h/J$ \cite{Johnston_1, Johnston_2}, which would imply in the quantum-mechanical problem that the Hamiltonian $\mathcal{H}$ has a gapless spectrum, and that certain correlation functions exhibit power-law decay in time \cite{Henley, Castelnovo_Chamon}, while remaining spatially short-ranged. We leave an exploration of this exotic possibility to future work. 
\acknowledgments
We thank Liang Fu for helpful discussions and comments.   This work is supported in part by DoE Office of Basic Energy Sciences, Division of Materials Sciences and Engineering under Award de-sc0010526, and by the National Science Foundation under Grant No. NSF PHY11-25915.

\appendix

\section{}
In the Appendix, we discuss the topological degeneracy of the $Z_{N}$ X-cube models on the three-torus.  We begin by elaborating on the degeneracy of the X-cube phase introduced in Ref. \cite{Fracton}, as understood from our isotropic layer construction.

The topological degeneracy of the X-cube fracton topological phase -- which was previously obtained \cite{Fracton_Gauge_Theory} by counting independent constraints on the operators $\mathcal{O}_{\rb}$, $A_{\rb}^{(j)}$ on the three-torus using an algebraic representation of the solvable Hamiltonian \cite{Haah} -- may also be understood by counting the number of topological sectors of the decoupled, two-dimensional toric codes that are made equivalent after condensation of the composite flux.  A more extensive discussion of this counting, is presented in the Supplemental Material \cite{SM}.  %
For example, the vacuum sector of the completely decoupled theory is equivalent, after flux loop condensation, to the topological sector where a single Wilson line -- which creates two-dimensional $Z_{2}$ flux excitations at its ends -- wraps around all of the $xy$ planes in the $x$-direction $W^{(xy)}_{x}$ \emph{and} all of the $yz$ planes in the $z$-direction $W^{(yz)}_{z}$. We label this topological sector of the decoupled toric codes as $W^{(xy)}_{x}W^{(yz)}_{z}$, which is now equivalent to the vacuum sector after composite flux condensation.  By cyclically permuting the $x$, $y$ and $z$ indices and by taking products of the Wilson line operators, we obtain seven additional topological sectors of the decoupled theory that are equivalent to the vacuum sector in the flux condensed phase [we may explicitly enumerate these sectors, adopting the previous notation, as $W^{(yz)}_{y}W^{(xz)}_{x}$, $W^{(xz)}_{z}W^{(xy)}_{y}$, $W^{(xy)}_{x}W^{(yz)}_{z}W^{(yz)}_{y}W^{(xz)}_{x}$, $W^{(xy)}_{x}W^{(yz)}_{z}W^{(xz)}_{z}W^{(xy)}_{y}$,  $W^{(yz)}_{y}W^{(xz)}_{x}W^{(xz)}_{z}W^{(xy)}_{y}$, and  $W^{(xy)}_{x}W^{(yz)}_{z}W^{(yz)}_{y}W^{(xz)}_{x}W^{(xz)}_{z}W^{(xy)}_{y}$].  
Since eight topological sectors of the decoupled theory are identified after condensation, we are led to conclude that on an $L_{x}\times L_{y}\times L_{z}$ three-torus, the X-cube fracton phase has topological degeneracy $D = 4^{L_{x} + L_{y} + L_{z}}/8$ or $\log_{2}D =2(L_{x} + L_{y} + L_{z}) - 3$ which reduces to the previously known degeneracy on the length-$L$ three-torus when $L$ is odd \cite{Fracton_Gauge_Theory}.

\begin{figure}
 \includegraphics[trim = 0 0 0 0, clip = true, width=0.2\textwidth, angle = 0.]{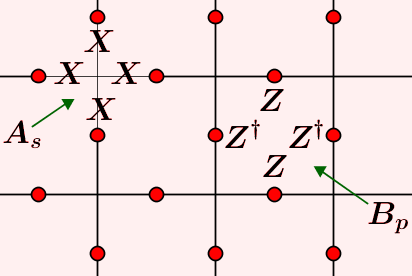} 
  \caption{{\bf 2D $Z_{N}$ Charge and Flux Operators}: Show are the star $A_{s}$ and plaquette $B_{p}$ operators that measure the $Z_{N}$ charge and flux, respectively, in the 2D $Z_{N}$ toric code.}
  \label{fig:Z_N_Toric}
\end{figure}

We may also compute the topological degeneracy of the $Z_{p}$ generalization of the X-cube phase, which we have introduced in the main text, when $p$ is prime. This fracton phase is constructed in the main text by coupling together copies of the $Z_{p}$ toric code, whose charge and flux operators are shown in Fig. \ref{fig:Z_N_Toric}.

For the $Z_{p}$ generalization of the X-cube phase, we find that the degeneracy on the length-$L$ three-torus is given by $\log_{p} D = 6L - 3$ when $L = p^{n} - 1$, through an extension of the arguments presented in Ref. \cite{Fracton}. First, the algebraic representation \cite{Haah, Polynomial} of the $Z_{N}$ generalization of the X-cube model is described by a stabilizer map $S = S_{x} \oplus S_{z}$ where
\begin{align}
&S_{x} = \left(\begin{array}{ccc}
-1 + \bar{x} & 1 - \bar{x} & 0\\
1 - \bar{y} & 0 & 1 - \bar{y}\\
0 & -1+\bar{z} & -1+\bar{z}
\end{array} \right)
\end{align}
and
\begin{align}
&\hspace{.2in}S_{z} = \left(\begin{array}{c}
-(1-z)(1-y)\\
-(1-x)(1-z)\\
-(1-y)(1-x)
\end{array} \right).
\end{align}
Here, $\bar{x} \equiv x^{-1}$, $\bar{y} \equiv y^{-1}$, $\bar{z} \equiv z^{-1}$, and the polynomials shown are elements of the Laurent polynomial ring $\mathbb{F}_{p}[x, \bar{x}, y, \bar{y}, z, \bar{z}]$ over the field $\mathbb{F}_{p}$ with prime $p$.  Let $R$ be the quotient ring $R \equiv \mathbb{F}_{p}[x,y,z]/\langle x^{L}-1, y^{L}-1, z^{L}-1\rangle$.   The degeneracy of the $Z_{p}$ X-cube model on the length-$L$ three-torus is given by $\log_{p}D = k_{x} + k_{z}$ where the quantities $k_{x}$ and $k_{z}$ are given by
\begin{align}
k_{x} &= \mathrm{dim}_{\mathbb{F}_{p}}\left[ \frac{R}{\langle (1-z)(1-y), (1-x)(1-z), (1-y)(1-x)\rangle} \right]\nonumber
\end{align}
and
\begin{align}
k_{z} &= \mathrm{dim}_{\mathbb{F}_{p}}\left[ \frac{R^{2}}{\left(\begin{array}{ccc} -1+x & 1-y & 0\\
1-x & 0 & -1+z
\end{array}\right)} \right]\nonumber
\end{align}

In the second expression, we have taken advantage of the fact that only two of the columns of $S_{x}$ are linearly independent. Both quantities $k_{x}$ and $k_{z}$ may be evaluated in the algebraic closure of $\mathbb{F}_{p}$, which we denote $\mathbb{F}$, when the length $L = p^{n} - 1$, so that $x^{L} - 1$ has $L$ distinct roots.  In this case, we determine $k_{x} = 3L-2$ by localizing at the maximal ideal $\langle x - t, \,y - 1, \,z - 1\rangle$, where $t\ne 1$ is a root of $x^{L} - 1$.  Similarly, $k_{z}$ is determined from the fact that the second determinantal ideal of $\left(\begin{array}{ccc} -1+x & 1-y & 0\\
1-x & 0 & -1+z
\end{array}\right)$ is precisely $\langle (1-z)(1-y), (1-x)(1-z), (1-y)(1-x)\rangle$.  As a result, localizing again at the maximal ideal $\langle x - t, \,y - 1, \,z - 1\rangle$ yields $k_{z} = 3L-1$.  This yields the desired result for the topological degeneracy of the $Z_{p}$ generalization of the X-cube model on a three-torus of length $L = p^{n}-1$.

\end{document}